%% file: main.tex
\documentclass[sigconf]{acmart}

\AtBeginDocument{%
  }

\copyrightyear{2025}
\acmYear{2025}
\setcopyright{cc}

\acmConference[CHI '25]{CHI Conference on Human Factors in Computing Systems}{April 26-May 1, 2025}{Yokohama, Japan}
\acmBooktitle{CHI Conference on Human Factors in Computing Systems (CHI '25), April 26-May 1, 2025, Yokohama, Japan}\acmDOI{10.1145/3706598.3714102}
\acmISBN{979-8-4007-1394-1/25/04}

\newcommand{\edits}[1]{\textcolor{black}{#1}}

\begin{document}


\title{Why So Serious? Exploring Timely Humorous Comments in AAC Through AI-Powered Interfaces}

\author{Tobias Weinberg}
\affiliation{
\department{Computer Science}
  \institution{Cornell Tech}
  \city{New York}
  \country{USA}}
\email{tmw88@cornell.edu}
\author{Kowe Kadoma}
\affiliation{%
\department{Information Science}
  \institution{Cornell Tech}
  \city{New York}
  \country{USA}}
\email{kk696@cornell.edu}

\author{Ricardo E. Gonzalez Penuela}
\affiliation{%
\department{Information Science}
  \institution{Cornell Tech}
  \city{New York}
  \country{USA}}
\email{reg258@cornell.edu}

\author{Stephanie Valencia}
\affiliation{%
\department{College of Information }
  \institution{University of Maryland}
  \city{College Park, MD}
  \country{USA}}
\email{sval@umd.edu}

\author{Thijs Roumen}
\affiliation{%
\department{Information Science}
  \institution{Cornell Tech}
  \city{New York}
  \country{USA}}
\email{thijs.roumen@cornell.edu}


\renewcommand{\shortauthors}{Weinberg et al.}

\begin{abstract}

People with disabilities that affect their speech may use speech-generating devices (SGD), commonly referred to as Augmentative and Alternative Communication (AAC) technology. This technology enables practical conversation; however, delivering expressive and timely comments remains challenging. This paper explores how to extend AAC technology to support a subset of humorous expressions: delivering timely humorous comments -witty remarks- through AI-powered interfaces. To understand the role of humor in AAC and the challenges and experiences of delivering humor with AAC, we conducted seven qualitative interviews with AAC users. Based on these insights and the lead author's firsthand experience as an AAC user, we designed four AI-powered interfaces to assist in delivering well-timed humorous comments during ongoing conversations.
Our user study with five AAC users found that when timing is critical (e.g., delivering a humorous comment), AAC users are willing to trade agency for efficiency—contrasting prior research where they hesitated to delegate decision-making to AI.  We conclude by discussing the trade-off between agency and efficiency in AI-powered interfaces, how AI can shape user intentions, and offer design recommendations for AI-powered AAC interfaces.
 See our project and demo at: \color{orange}{\href{https://tobiwg.github.io/research/why_so_serious}{tobiwg.github.io/research/why\_so\_serious}}
\end{abstract}

\fancyfoot{}


\begin{teaserfigure}
  \includegraphics[width=\textwidth, alt={We explore how to design interfaces to support timely humorous comments in Augmentative and Alternative Communication (AAC). On the left side there is (a) an AAC user tries to make a joke in an ongoing conversation. On the right in the upper half (b) as the conversation continues,
the user frantically tries to type the joke. On the right most in the upper half (c) unfortunately they trigger text-to-speech (TTS) too late to effectively land the joke causing
Confusion. On the bottom right (d) we explore how to design interfaces to speed up the response through AI powered tools, while minimizing the cost of
agency. On the right most of the bottom(e) Leading to design that allows AAC users to meaningfully express themselves through humor in ongoing conversations.
Note: AAC input types and interfaces vary widely across users. One of our authors is an AAC user who commonly interacts with his
phone as displayed here. Only for illustration purposes. Our actual interfaces were used on a range of devices in our studies}]{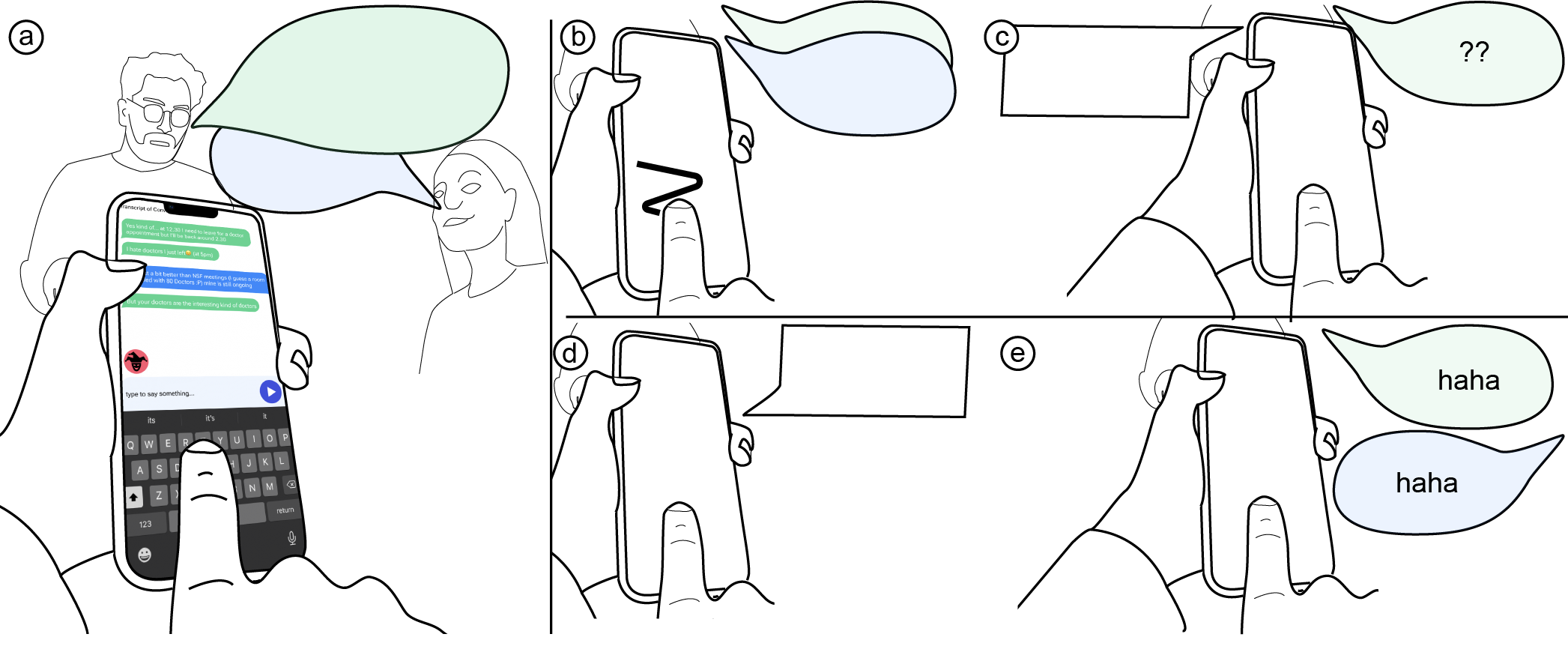}
  \caption{We explore designing AAC interfaces for timely humorous comments. (a) An AAC user attempts a joke. (b) They frantically type as the conversation moves on. (c) The joke plays too late, causing confusion. (d) We explore AI-powered tools to speed up responses while preserving agency. (e) Our designs help AAC users express humor meaningfully Note: AAC input methods vary. The lead author, an AAC user, interacts with their phone as shown. Our study tested interfaces across devices.  }
  \label{fig:teaser}
\end{teaserfigure}

\makeatletter
\def\@ACM@copyright@check@cc{}
\makeatother
\maketitle

\input{src/00_introduction}

\input{src/01_relatedwork}
\input{src/02_Interviews}

\input{src/03_system}

\input{src/04_evaluation}

\input{src/05_futurework}

\input{src/06_conclusion}

\begin{acks}
We extend our sincere gratitude to Zekai Shen, Samee Chandra, and Maria Tane for their invaluable contributions to the development of the prototypes. We also thank Roy Zunder for his assistance in data analysis and writing. This work would not have been possible without your support and dedication.
\end{acks}

\bibliographystyle{ACM-Reference-Format}
\bibliography{CHI2024}


\end{document}

%% file: src/00_introduction.tex
\section{Introduction}

Augmentative and Alternative Communication (AAC)  devices are vital for people with disabilities that affect their speech because they enable them to communicate effectively. In combination with accessible input devices such as eye tracking, switches, pointers, and customized vocabulary layouts, AAC devices support users with a wide range of cognitive, motor, and verbal abilities to communicate. While current AAC devices work for practical everyday conversations, they pose challenges for time-sensitive or expressive forms of communication, like making timely humorous comments. 

Researchers have explored the challenges of communication asymmetry in AAC and strategies to mitigate its impact. For example,
\citet{waller_telling_2019} found that AAC users communicate at approximately 12-18 words per minute (compared to ~125–185 wpm for typical users), resulting in what is known as \textit{communication asymmetry}. Researchers have attempted to overcome this asymmetry through word prediction~\cite{trnka_user_2009}, visuals~\cite{fontana_de_vargas_aac_2022, fontana_de_vargas_automated_2021}, and by inferring word suggestions from context, such as location, conversation transcripts, and image recognition from the surroundings \cite{kane_lets_2017, kane_what_2012,valencia2024compa}.

In clinical contexts, the challenge of speeding up AAC programming (e.g., utterances or messages) has been widely studied too \cite{holyfield2019programing, schlosser2016just}. The concept of Just In Time (JIT) programming of AAC devices provides users with contextually relevant speech suggestions \cite{holyfield2019programing}. We build on the concept introduced by \citet{holyfield2019programing} by integrating Large Language Models (LLMs) and automatic speech recognition ~\cite{wisenburn2008aac,valencia2024compa}. These technologies help AAC users navigate the fast-paced and unpredictable nature of informal conversations, particularly when trying to make humorous comments.

Researchers and practitioners have shown a growing interest in incorporating LLMs to enhance auto-complete and suggestions features to improve AAC users' communication speed \cite{shen_kwickchat_2022, valencia_less_2023}. Recent work has focused not only on enhancing typing speed but also increasing opportunities for AAC users to respond promptly, especially in the presence of communication asymmetries. For example, \textit{COMPA}~\cite{valencia2024compa} enabled AAC users to mark conversation segments they intend to address and generate contextual starter phrases based on the marked conversation segment and the selected user intent. This proved to be an excellent tool to support both AAC and non-AAC users to have a better conversation flow. 
Similarly, \textit{KWickChat} ~\cite{shen_kwickchat_2022}, a sentence-based text entry system that automatically generates sentences based on keywords. Using \textit{KWickChat}, AAC users generated meaningful replies in daily conversations with very little user input. Although these systems improved the communication rate of AAC users in specific settings, we observe a risk that excessive reliance on AI-generated auto-complete features may limit users' ability to express less predictable and spontaneous comments through AAC technology. Additionally, while supporting faster speaking rates is important, communication and AAC use extend beyond transactional conversations. AAC users, like everyone else, are interested in diverse and rich interactions such as storytelling, telling jokes, and sharing experiences~\cite{waller2006communication}.

In this work, we study the trade-offs between efficiency vs agency of AI-assisted communication for AAC users. We built interfaces that leverage AI at different levels of automation to address one of the most challenging facets of expressive communication: timely humorous comments during ongoing conversations. Humor helps manage the flow of a conversation and helps build common ground between participants, making interactions enjoyable and fostering interpersonal connections~\cite{Waller2009STANDUP, binsted2006computational, omara2006designing}. Early work on humor interfaces for children using AAC explored algorithms for pun generation~\cite{omara2006designing}. This work demonstrated the value of wordplay for expressive humor creation. Using today's LLMs we explore different interactions to support AAC users in timely humorous comments during an ongoing conversation.

In this paper, we investigate how to better design AAC technology to support making \textit{timely humorous comments}. Creating humorous interjections poses significant challenges due to the need for both timely responses and expressive agency. We focused on this challenging use case because it requires precise timing and tends to be unpredictable. We extend AAC technology by designing interfaces powered by multi-modal LLMs, to facilitate this form of social interaction. By utilizing real-time transcription, analyzing conversation context, extracting keywords, and generating speech output, we harness the capabilities of multi-modal LLMs such as GPT4~\cite{openai_gpt-4_2023} and LLAMA3 to interact with the context of the conversation.

We explore this use case for AAC through three studies. (1)~First, to understand the current usage and challenges of AAC technology and the role humor plays in the lives of AAC users, we conducted seven qualitative interviews with people who use AAC. (2)~Second, we take a design-driven approach to explore the trade-offs between different design considerations resulting in four interfaces to create humorous comments with AAC technology. (3)~Finally, to understand how our design decisions impact users' experience, we conducted a user study with five AAC user participants. We conclude by discussing the trade-off between agency and efficiency in AI-powered AAC interfaces, and how AI-generated suggestions can influence user intentions and communication styles. Additionally, we provide design recommendations to ensure that AI-powered AAC interfaces maintain user agency while supporting efficient and expressive communication.

\section{Positionality Statement}
This work is uniquely positioned not just in terms of the problem we attempt to tackle but also the personal background of the lead author of this work. The lead author of this work has a speech disability and is a user of AAC technology. To understand how this impacts the positioning of the work, we disclose his perspective as it plays a critical role in how some of the work was conducted.

The lead author lost his ability to speak when he was 15 years old, \emph{``it was a wild experience see my whole world changing in just one month when the neuro-motor disease manifested, suddenly, I was not able to speak or move normally''} this lived experience made him determined to help people with similar disabilities. \emph{``Humor is the lens through which I see the world. It is a part of me I can't help it. When I lost the ability to speak, I had to learn to adapt my humor to this new form of communication without losing my touch''.}

The lead author entered the AAC world without knowing what AAC was. \emph{``Growing up in Argentina, the access to technology is a bit different. No one told me ``here, these are AAC devices you can use''. I had to discover everything on my own''}. At first, while he was in the hospital, he used a laptop to type what he wanted to communicate \emph{``I never felt that this condition was a limitation in my life. It was just a new way to experience life, and it was not going to stop me''.}

Initially, he used an iPad mini, but the TTS lacked expressiveness. \emph{``The voice was monotone and without any emotion. I never felt identified with it. It was very hard to convey what I wanted and how I wanted it. The deal breaker for me was the accent, as it only had Mexican- or Spain-Spanish voices. I didn't like sounding different from all my friends. I don't think anyone made fun of this, but internally I felt like an outsider''}. Eventually, he switched to using the notes app on an iPhone and showing others typed text \emph{``I found that showing the text was better for me because when people read, they automatically put the `correct' tone in their heads. That, paired with my facial expression while people read my messages, allows me to express myself much better. Now, I don't have any specific accent I have the accent that the reader imagines in their head''.}

He has now been an AAC user for over 11 years, and his experience is catalyzed in this work \emph{``I feel that everyone should be able to express humor how and when they want. Humor is an extremely powerful tool to connect with people. It is easy to take for granted, but for us (AAC users), making timely humorous comments is not trivial. I am very lucky to be able to type fast. However, even I often find myself in situations where the joke window is long gone by the time I am done typing, and my humorous comment is no longer relevant. This inspired me to set out to study how to build AAC technology to better express ourselves.''}

\section{Contribution, Benefits, and Limitations}

We explore how to extend AAC technology to enable more expressive speech. In doing so, we focus on a challenging form of communication that benefits from the precise use of words,  and context: Timely humorous comments. We contribute insights from three research activities:

\begin{enumerate}
  
    \item From in-depth interviews with seven AAC users, we gain insights into the current usage and limitations of AAC technology for expressive speech and the role humor plays in the lived experience of our users.
    \item We explore a design space of interfaces for AI-powered apps to support AAC users in making timely humorous comments. We discuss trade-offs in design choices, specifically around user control and efficiency.
    \item Through a user study with five AAC users, we evaluate four proposed interfaces to derive insights on interface design and identify preferences and trade-offs regarding the identified design space.
    \item Finally, based on the conclusions of these three activities, we draw design recommendations for the development of future AAC technology.
\end{enumerate}

We do not claim generalizability of our conclusions, as is common in the field of accessibility; we had a small but very diverse set of participants for our studies. The conclusions are built on the insights and experiences from our group of AAC users, but we do not claim full coverage of people who use AAC. The prototypes addressed a subset of the interview insights and the first author's personal experience. Nevertheless, they do not cover all aspects discussed in the interviews. However, insights from these interviews will inspire future work in this area. The design space we explored specifically focuses on components in user interfaces of AI-powered applications to create timely humorous comments. We consider this an important and emerging subset of the space. Still, other aspects are worth exploring, such as different input technologies or giving users control over the actual voice for expressive speech. Finally, we have studied the usage of these interfaces in a relatively controlled environment. Our conclusions will inform the design of interfaces that would be worth testing in more open-ended scenarios or long-term deployment.

%% file: src/01_relatedwork.tex
\section{Related Work}

We build on work on challenges in AAC and expressiveness, trade-offs in AI writing support, computational humor, as well as existing practices in AAC comedy.

\subsection{Challenges in AAC: Timing, Agency, and Expression}

AAC users communicate at approximately 12-18 words per minute (wpm) compared to 125–185 wpm for non-AAC speakers~\cite{waller_telling_2019}, and in fast-paced conversation, this may often cause them to miss opportunities to participate. \citet{higginbotham2016time} studied conversations between AAC users and their interlocutors, finding that the perception of an AAC user’s competence is influenced by how well their conversational timing aligns with their partner's. Similarly, \citet{higginbotham2013slipping} showed how mismatched timing in conversations disrupts the flow and creates barriers to effective communication. These findings highlight the critical role of timing in shaping conversational dynamics and perceptions of AAC users' capabilities. Motivated by these findings, we contribute to interaction and timing for AAC users by studying how to support timely humorous interjections.

Besides timing, \textit{conversational agency} (an individual's capacity to express and achieve their goals in conversation) is influenced by social constraints~\cite{valencia_conversational_2020}, the pace of communication~\cite{kane_at_2017}, and the quality of communication resources available (e.g., vocabulary, technology, physical and social context)~\cite{ibrahim2024common, curtis2022state}. In building tools that impact agency, it is essential to prioritize empowering the user's intentions while avoiding overreliance on purely automated content~\cite{valencia_less_2023}.

We are not the first to address existing asymmetries in timing and agency across AAC users and their interlocutors; researchers have increased AAC users' communication rates by incorporating LLM-based suggestions~ \cite{shen_kwickchat_2022, valencia_less_2023, yusufali2023bridging, cai_using_2023, fang_socializechat_2023}, using visual output, or even physical attributes ~\cite{bircanin2019challenges, curtis2024breaking} to support the conversational flow~\cite{valencia2024compa,sobel2017exploring,fiannaca2017aacrobat}. We put these ideas to the test in the high-paced context of delivering humorous comments.

In clinical contexts, the role of timing for AAC is a well-studied phenomenon. \citet{schlosser2016just} introduced the notion of \textit{Just-In-Time~(JIT)} vocabulary programming, adopted from a classical business context in clinical AAC. Using JIT, clinicians can manually switch the content based on changing contexts which stands in stark contrast to preemptively programming all contexts. Recent work shows that JIT can also benefit language learning~\cite{holyfield2019programing}. We build on this notion by leveraging AI to present contextually relevant content in AAC interfaces in near real-time.

\subsection{Trade-offs in AI Writing Support}
Outside of AAC technology, the goal of providing tools to auto-complete or generate written text has been around for a long time. Initially, text entry systems focused on a word~\cite{bi2014complete} or short sentences~\cite{vertanen2015velocitap} based on likelihood distributions. More recent systems, however, suggest multiple short replies~\cite{kannan2016smartreply} or a single long reply. This unlocked new use cases for LLMs such as inspiring ideas~\cite{gero2022sparks}, revising text~\cite{cui2020text}, and generating stories~\cite{singh2023elephant}. It is not surprising that there is an interest in viewing AI-writing assistants as co-authors~\cite{lee2022coauthor,mirowski2023screenplays,jakesch2023opinion, Goodman_2022}. However, similar to the agency considerations,  these tools raise concerns regarding their impact on the user’s authenticity and agency. AI-assisted writing environments may hinder individual uniqueness and authenticity \citet{kadoma2024role} study this phenomenon in the workplace and its effect on inclusion, control, and ownership when collaborating with LLMs.

Similarly, \citet{valencia_conversational_2020} suggest that AAC users are concerned about social perceptions when using AI-generated text. They found that AAC users worry that interlocutors may perceive them as lacking the ability to communicate or as unwilling to put effort into communication if they rely on AI-generated responses. In designing AI interfaces for AAC users, we want to pay specific attention to their perceived agency and authenticity.

\subsection{Computational Humor and Creative Expression Tools}
The idea of creating tools for humorous expression has been studied in a field called Computational Humor for the past 3 decades~\cite{ritchie2001current, stock2002april, hulstijn1996proceedings}. Early systems in this field generated basic puns~\
\cite{lessard1992computational} and riddles~\citep{lessard1993computational}. A bigger project JAPE~\cite{binsted1994implemented, binsted1996machine, binsted1997computational} built on \textit{WordNet}~\cite{miller1990introduction, fellbaum1998towards} dramatically expanded the capabilities of such systems. Others explored other media to generate humor such as images~\cite{Wen2015OMGUF}. With the advent of LLMs, computational humor started to explore the interpretation of humor as well~\cite{rayyashAI}. Computational humor has shown that humorous AI systems, notably \textit{Witscript}, can appear human-like during joke improvisation~\cite{toplyn2023witscriptgeneratingimprovisedjokes}. In a study using \textit{Witscript}, a human, \textit{GPT-LOL} (a simple joke generator created to serve as a baseline), and \textit{Witscript~3} jokes were rated for human-likeness. \textit{Witscript~3}'s responses were judged as jokes 44.1\% of the time in comparison to \textit{GPT-LOL} 33.8\% and 23.6\% for human responses. 

Humor is not new in the context of AAC research. Notably the \emph{STANDUP} project~\cite{manurung2008construction, ritchie2007practical, Waller2009STANDUP, omara2006designing} enabled children with complex communication needs (CCN) to develop their linguistic skills through humor creation. The \emph{STANDUP} program (inspired by Binsted’s JAPE program) was a great showcase of the power of using humor to enrich the lives of children with CCN. While STANDUP acted as a great language playground for children enabling them to compose puns, we explore how different input modalities impact the perception of agency when interacting with AI assistance to create timely humorous comments during an ongoing conversation.

Similar to \cite{kantosalo2014isolation} we aim to better understand the roles that computers and AI agents can play in supporting humor expression for AAC users. While systems explored in prior work focus on creative expression in asynchronous settings~\cite{tamburro2022comic, welsh2018ticket}, in this paper we explore creative humor expression in real-time conversation, proposing different interaction techniques to alleviate timing constraints. 

\subsection{AAC \& Comedy}
Humor plays an important role in the (public) lives of existing AAC users. Some showcase comedic talent through pre-programmed messages, allowing for carefully crafted and timed humor in performances. Examples include comedians Lee Ridley (Lost Voice Guy), Nina G, and Ahren Belisle. Ridley, the 2018 Britain’s Got Talent winner, uses his AAC device to deliver sharp, pre-written jokes; Nina G, the world’s first female stuttering stand-up comedian, occasionally integrates AAC into her acts alongside her unique advocacy for disability awareness; and Ahren Belisle, a finalist on America's Got Talent in 2023, a comedian and engineer, employs his AAC device to deliver humor that reflects his experiences and sharp wit. While this highlights the agency AAC users already exercise in creating humor, our research focuses on a different challenge: enabling timely, humorous comments during dynamic, ongoing conversations. This context requires AAC technology to support rapid, contextually appropriate humor generation.

%% file: src/02_Interviews.tex
\section{Qualitative Interviews: understanding use and challenges of AAC technology to create humorous comments}

AAC technology has come a long way, and AAC users already incorporate humor as part of their daily lives. To better contextualize the role humor and AAC technology play in their lives, we conducted qualitative interviews with our target population. The interview protocol and data processing were reviewed by Cornell University's IRB and approved under protocol IRB\#0148755.

\subsection{Participants}
We recruited seven people with speech disabilities, ranging from diverse conditions; Cerebral Palsy, head traumatism due to an accident, neuromuscular condition, and autism spectrum. there was a diverse range of the severity of the motor disabilities we grouped them into the following categories:  Mild (minor limitations in movement), 
 Moderate (noticeable limitations, performs daily activities with some assistance),
 Profound (unable to perform daily activities without significant assistance), and
Severe (requires assistance for most daily activities).
\begin{table*}[t]
    \centering
    \resizebox{\textwidth}{!}{
    \begin{tabular}{llllll}
         \textbf{ID} & \textbf{Age} & \textbf{Gender} & \textbf{Motor Disability} & \textbf{AAC Device/s} & \textbf{Input Modality} \\
         P1 & 25-34 & Female & Mild  & Phone/iPad  & Direct touch \\
         P2 & 25-34 & Non-binary & Mild  & Laptop/iPad & Direct touch/keyboard \\
         P3 & 25-34 & Female & Profound  & Accent1400, Nuvoice & Direct touch with guard \\
         P4 & 45-54 & Male & Moderate & Accent 1000 & Direct select input \\
         P5 & 45-54 & Male & Profound  & Tobii Dynabox & Eye gaze \\
         P6 & 25-34 & Male & Profound & Eye gaze edge/grid 3 & Eye gaze/track-ball \\
         P7 & 35-44 & Male & Severe & Tobii Dynabox & Eye gaze \\
    \end{tabular}
    }
    \caption{Participant details}
    \label{tab:participants}
\end{table*}
Table~\ref{tab:participants} contains an overview of their basic demographics. To increase access to our study, we allowed participants to attend remotely via video conference software. This approach increased the diversity of participants we recruited.

\subsection{Structure}
We followed a semi-structured interview process. The two main themes we were interested in exploring were: (1)~challenges and usage of AAC for expressive communication, and (2)~the role humor plays in the lives of our participants. We include the full interview protocol in the supplementary material. 

As mentioned in the positioning statement, the lead author of this work has a speech disability. Thus, interviews were conducted using his AAC device, which added a layer of observation to the conversation and interview process.

\subsection{Data and Analysis}
We transcribed video and audio recordings of seven interviews. All interviews were conducted remotely through video conference software. These interviews lasted between one hour to three hours to accommodate each participant's needs and speech rate. 

We used thematic analysis \cite{braun} to derive insights from the interviews. Three members of the research team individually conducted open coding on the interview transcripts. First, they coded the same transcript individually. Then, they came together to review discrepancies to reach a consensus and split the remaining transcripts. After that, the three researchers conducted two rounds of affinity diagramming together to thematically group the codes. Finally, they converged their code books to a total of 62 codes and extracted three higher-order themes with six underlying sub-themes. For the complete code book, please refer to the supplementary materials.

\subsection{Results}
This section outlines the main themes illustrated by relevant quotes and observations. We identified the following overarching themes: reliance on AAC, current use and experience of humor with AAC, and barriers to humor expression. Each theme is further divided into more specific observations from the interviews, detailed in the sections below.

\subsubsection{Reliance on AAC: "Communication is wild and crazy, but it is also the window to my world"}

Participants shared various reflections on their AAC experiences, emphasizing its challenges and benefits. For all participants, AAC is essential for communication. P2 described AAC as "freedom," allowing them to avoid rationing speech: \emph{“I can often rely on mouth words, but not always, and having AAC available ... it means I don't have to worry about rationing speech.”} P3 echoed this sentiment, calling AAC a "blessing" that helps them stay connected: \emph{“I would be lost without it.”} Similarly, P1 shared how AAC has improved their interactions with loved ones: \emph{“[AAC technology] helped me communicate with friends and family. When I talk with my voice, and  they can’t understand me.”}

Humor is central to how participants interact and relate to others. P6 explained how humor lightens the burden of AAC use, making interactions more enjoyable and natural: \emph{“We laugh a lot. It makes using AAC less of a chore… and makes us like regular guys.”} P4 added that humor comes naturally to them: \emph{“I laugh a lot and find things funny”}. Yet, P5 highlighted the complexity of navigating others’ reactions to humor, noting that people sometimes laugh nervously because they are unsure how to respond: \emph{“When I am funny, people laugh in a nervous laugh, they don't know how to take me. I'm just like everyone else.”}

Participants shared how AAC opens up their world, giving them access to experiences like everyone else. P7 captured this sentiment, saying: \emph{“I learned this communication is wild and crazy, but it is also the window to my world, giving me the chance to do so many things just like everyone else.”} P6 agreed: \emph{“That is life as an AAC user—accept is good enough.”}

Overall, participants’ reflections show how AAC not only facilitates essential communication but also supports humor, emotional expression, and personal connection, despite its challenges.

 \subsubsection{ Current Use and Experience of Humor with AAC}

This recurring theme provided two contexts in which humor plays an important role for AAC users, each with different implications. We divide this theme into two categories: humor as an avenue for \textit{inter-personal} relationships, and humor for \textit{personal} expression.
 
\textbf{Humor as an avenue for inter-personal relationships.} Humor functions as a form of social capital for AAC users. Participants frequently discussed the significance of humor in their lives. P3, for example, emphasized humor’s central role in fostering connections: \emph{“[Humor is] very important. My family and I are always joking around. Staff and I too.”} then she elaborated \emph{"[Humor allows her to ] Interact with people better, Connecting better with people who I am talking to"} This reflects how humor fosters a sense of normalcy and closeness with others.

Humor can also be a means of contributing to conversations and integrating more fully into social interactions. P1 described a moment where humor enabled them to engage in a conversation: \emph{“When my friend is saying something funny, then I join in by adding a few words to help them out”}. In doing so, they illustrated humor’s capacity to bridge the gap between AAC users and their conversational partners. P1 further elaborated: \emph{“Sometimes I make everyone laugh because it allows everyone to enjoy my jokes”}, demonstrating how humor allows AAC users to shift the dynamic and become a focal point of joy and connection.

Several participants shared humor’s role in making them feel more relatable and human. P6 explained how humor helps break down barriers in conversation, whether casual or formal: \emph{“Humor is important in conversations for SGD users, just like any other person who can use their physiological voice. For example, in casual conversation or when giving a speech, you want to get the audience to relax and relate to you more.”} Similarly, P7 reflected on the importance of humor in being recognized as a person, adding that it helps \emph{“to get people to see that I am a person.”} P4 shared a related example during a webinar about self-advocacy, where humor helped them connect with their audience: \emph{“I was saying because I am an old fogey that I had more experience than others—people laughed at that.”} This illustrates how humor can defuse tension or build rapport, even in more structured settings.

These reflections reveal that humor is not just a tool for interaction but also enhances AAC users' sense of self-worth and social integration. P6 succinctly captured this sentiment: \emph{“Humor is important. Using an AAC is hard enough, and I love joking around with my friends.”} P5 echoed this by adding, \emph{"I like to use humor every day. I like to laugh; I always have."} Together, these statements underline the critical role humor plays in helping AAC users connect, relate, and seamlessly integrate into social environments—acting as a crucial form of social capital.

\textbf{Humor as an avenue for personal expression.} We observed that humor is often personal and subjective, and the way AAC users express humor revealed diverse preferences. For example, P1 mentioned using common phrases like \emph{“Oops, I did it again,” or “How rude?”} to convey expressive reactions during a conversation. P4 shared, \emph{“For me, it's just when I find something funny or when I'm writing a speech I can naturally put things in that I think are funny, not necessarily that they are funny to other people.”} These examples illustrate how humor is often personal and subjective.

Humor is understood and used by AAC users in varied ways.
P3 emphasized the playfulness in the language, describing their preference for \emph{“play on words, not Shakespearean style, but stuff to joke around, Sarcasm, Light conversations.”}. P2 expanded \emph{"I definitely use a lot of humor like saying true things in humorous ways that point out the absurdity So, satire-like but not quite satire? Not quite sarcasm either, usually. Like saying true things in ways that point out the absurdity."} P4 and P5 added that they use sarcasm a lot.

P6 highlighted that AAC should include genres of jokes, such as \emph{“your mama jokes”}. They furthermore explained, \emph{“Having the most current popular expressions, including single words, and having a speech device that updates regularly to add them. It shouldn't take any away because you might want to continue using a particular word or saying, even if it isn’t the most current.”} This illustrates the dynamic nature of humor and how language evolves over time.

AAC users express humor in varied ways, highlighting the complexity of designing interfaces that can capture this diversity in styles and showing how humor preferences can be culturally and contextually specific.

\subsubsection{Barriers to Humor Expression}
We found three major barriers for AAC users that hinder their ability to freely express their humor: timing, intonation, and technical limitations. Here, we provide examples of our participants' lived experiences and perspectives to explain how these factors affect their contextual communication choices and how they express humorous comments.

\textbf{Timing: “Timing is everything.”} We asked participants what they prioritize when delivering humorous comments: timing or expressiveness. It was a surprisingly even split; a slight majority prioritized expressiveness (P2, P4, P6, and P7) over timing (P1, P3, and P5); they clarified they have to gauge their conversational partner expectations to decide how to communicate moment to moment. Related, and maybe more importantly, P1, P3, P4, P5, P6, and P7 agreed that the context window where humorous comments can land is short, and they feel they are always \emph{“falling behind”} in conversations. P4, P5, P6, and P7 shared a similar feeling,  P4 exemplified: \emph{“Sometimes I see and hear something funny while I'm typing something else mid-sentence. I can't express that and keep what I was saying”}. 

Participants mentioned strategies to deliver timely humorous comments: (1)~Signaling intention to speak (P2), (2)~using expressive reactions (P4), (3)~shortening or simplifying comments (P1, P3), and (4)~waiting for their turn to speak (P1). Strategies 1 and 2 provide AAC users control over timing, while strategies 3 and 4 respond to the conversational partner's control of the conversation. P1 explained  \emph{“Sometimes you have to sacrifice what you would want to say because it's too long”}. P7 added that they allow people to continue speaking while they are typing because they feel \emph{“the slowness gives anxiety (...), and people cannot keep up with the conversation”}. 

Participants explained how they leverage pre-generated messages to deliver their humorous comments faster. Some participants, like P1, P2, P3, P5, and P6, were accustomed to generating phrases and then using them at a moment’s notice. They explained they use a built-in functionality in their SGD to write down predefined phrases like “I am sorry about that” or “ha!” (P6) to react humorously to their conversational partner commentary. While these phrases can be used as humorous reactions during any conversation, by definition, these phrases are context-agnostic and do not reflect any deeper thoughts or direct references to what is being said during the conversation. 

To summarize, timing is a major barrier for AAC users when making humorous comments. Providing ways for AAC users to deliver contextually appropriate humorous comments faster while retaining their original intention remains an open question.

\textbf{Intonation: It’s not what you say but how you say it.} A common challenge AAC users faced when creating humorous comments was a lack of control over the intonation of their device (P2, P3, P4, P5, P6, and P7). Participants felt that these devices did not reflect the tone of their \emph{“inner voice”} to deliver jokes appropriately. For example, P4 explained how the voice does not match how they think they sound: \emph{“The tone I have in my head don't translate well to the device. It may seem angry and not funny when I say it with the device… We get frustrated because people don't hear it the way we hear it in our head”}. Still, P4 emphasized that with practice, they learned to adapt their timing and cadence: \emph{“The more I get the feel of my timing with the device... you can put jokes in your cadence because you hear it and start speaking in your head like the device.”}

Thus, participants felt that tone was a fundamental aspect of communicating humorous comments and \emph{“sounding like a human”}, equal to people who can speak with their own voice. P6 said that the thing that they dislike the most about their AAC device is that \emph{“There isn't much voice inflection, or the ability to regulate the speed without going into settings”} and that to tell a joke \emph{“you might want to slow down part of what you are saying, like a pre-emphasis”}, explaining how tone and speech rate could enhance the delivery of the joke. P5 added that \emph{“sarcasm is very hard to convey as an AAC user”} because \emph{“this device doesn't have emotion”}. P3 further detailed that the device makes her sound monotone since she cannot change the tone \emph{“right away”}. This impacts how she thinks others perceive her. Lack of control over intonation is why P2 sometimes prefers to project their speech with a projector, similar to the lead author's point, if the words they want to say are read, the tone comes across better. P2 explained: \emph{“Flat tone of voice can be a problem when I'm using speech generation ... but I usually don't do that anyways(...) TTS is so bad. Like, it's better than not having the option. But that's different from being good.”}

All participants use a combination of non-verbal gestures (facial expressions, hand gestures, and bodily movements), and their natural voice to react to ongoing conversations. For example, P4 detailed how they switch between their natural voice and their device for expressive preference: \emph{“Expressive reactions come from my natural voice and I use my device for clarity for people who don't understand me”}. Non-verbal gestures play an important role in signaling humor: \emph{“it's usually the looks I give or the sounds I make that are letting people know I'm joking"}. However different AAC users have different expressive capabilities. P1, P2, and P4 can use natural speech and move their bodies to express themselves, while P6 cannot move their body on their own. 

Overall, participants' lived experiences show how they cope with the lack of intonation in their devices, but these techniques are not universally applicable. This creates significant barriers to conveying humor and nuance, which, as discussed before, play a fundamental role in AAC users' communication choices and self-perception.

\textbf{Technical Limitations of AAC.} Participants brought up technical challenges of AAC technology hindering their ability to make humorous comments. Two limitations directly affect AAC users’ expressiveness: Lack of multilingual support and correct pronunciation of words in AAC Software. These problems were highlighted by P2 and P6. P2 is a bilingual speaker of English and Mandarin, and they expressed that: \emph{"Mandarin Chinese AAC's inability to detect which pronunciation should be used for characters pronounced differently based on context is 100\% annoying.”} This problem was not unique to multi-lingual users, P6 clarified that English-language AAC has a similar problem: \emph{“I hate that it does not pronounce words that are spelled the same correctly, like read, or lead”}. 

As for limitations that indirectly affect users’ expressiveness, common challenges were battery life and Internet access. Participants mentioned they rely on AAC technology to communicate with others, so they chose software to guarantee their devices were always available. P3, P4, P6, and P7 strongly preferred software that doesn't rely on internet access. They adopt similar patterns to avoid battery drain, P4 explained: \emph{“Unless I'm home, I don’t turn on my internet because it drains batteries faster”}.  

Finally, participants mentioned indirect barriers posed by proprietary AAC software and hardware. AAC users' range of mobility varies greatly, and this is reflected in the wide variety of their software and hardware choices. For instance, P2 uses at least 5 different applications to communicate in different settings (e.g., Catalystwo, Proloquo4Text, Flip Writer, projecting text editing software, and eSpeak), and P3, P4, P5, and P6 all use different devices with different input assistance (e.g., eye tracking, Accent 1000, Nuvoice, etc.)

%% file: src/03_system.tex
\section{Exploring a design space of AAC interface design to create  humorous comments}

By exploring a part of the design space of AAC interfaces, we aim to better understand the trade-offs involved in designing such applications. Based on the critical role timing plays in AAC devices, as seen in the qualitative interviews (5.4.3 ``timing is everything'') and the lead author's first-hand experience, we decided to focus on techniques that leverage AI as a form of autocomplete to create timely humorous comments.
While voice and intonation also emerged as important themes in delivering humorous comments, these aspects fall outside the scope of our prototypes. Prior work has made significant strides in incorporating tonal nuances into AAC~\cite{pullin201517of, pullin2017designing, aylett2016don, preece2024making, fiannaca2018voicesetting}, and we see potential for generative AI voices, like the work of \textit{ElevenLabs}, to help AAC users customize their own unique voices.

While auto-complete will enable to make humorous comments faster, there is a risk that it diminishes the user’s sense of agency by making jokes \textit{for users} instead of \textit{with users}. To explore this tension, we consider different ways LLMs could suggest contextually relevant, humorous comments in real-time and implement four prototypical interfaces supporting different degrees of agency (user control) and efficiency (time to deliver the humorous comment).

\subsection{An Example Interface}

\begin{figure}[h]
  \centering
  \includegraphics[width=0.33\textwidth, alt={Overview of an example AAC interface to create humorous comments (\emph{Wizard} interface). (a) We extended this app with a ``joke mode'', which triggers the different forms of interfaces we explore in this section to support users in composing humorous comments. (b) the user presses this button to send the joke to be read by the tts (c) This example interface, later referred to as \emph{Keywords}, displays keywords
based on the ongoing conversation, which is captured and transcribed using automatic speech recognition (ASR). (d) Users either start typing their joke and/or select keywords to compose it. (e) The interface displays auto-completed suggestions for jokes and updates these as users provide more context. (f) If the user does not like the suggestion they can hit the refresh button to get a new one.}]{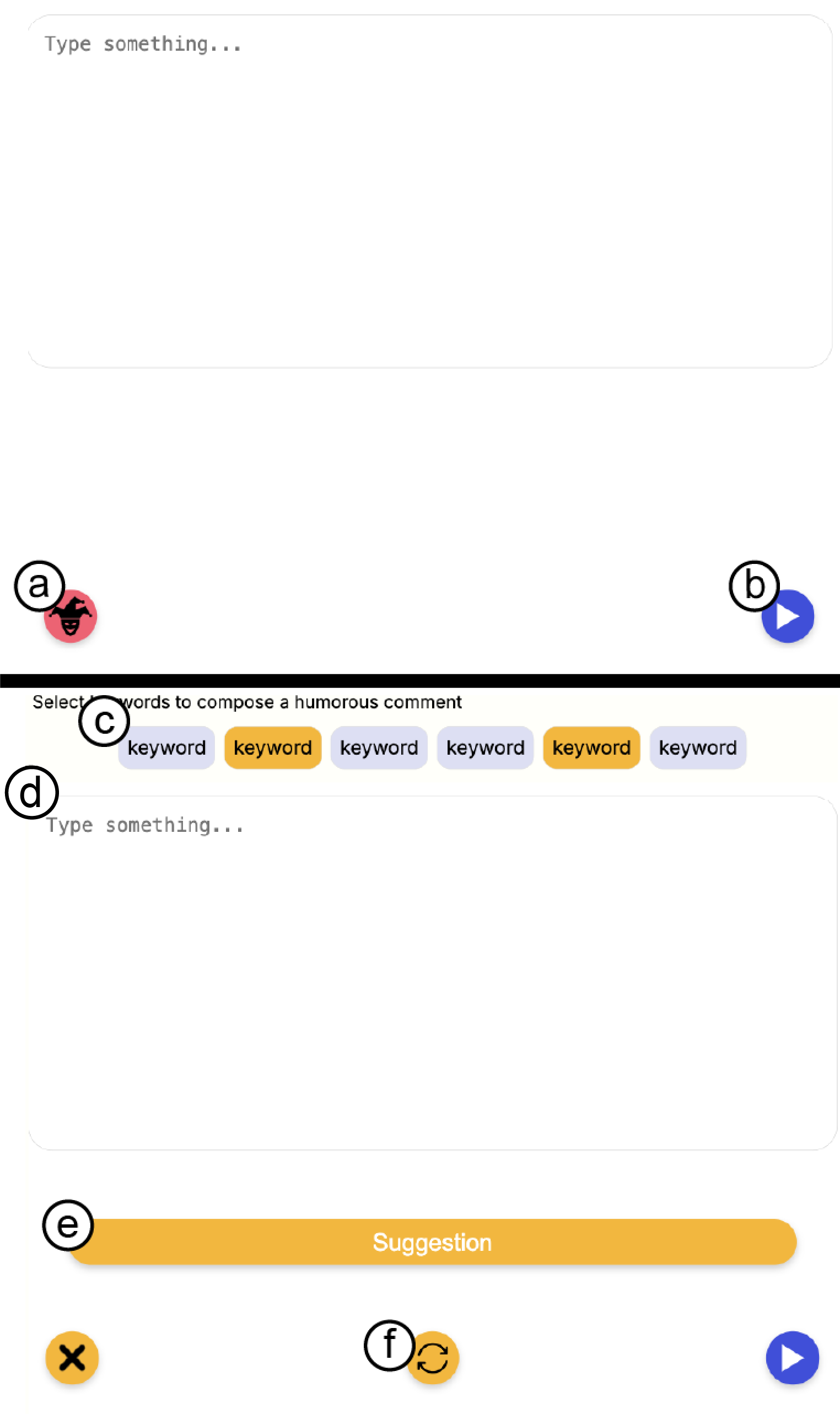}
    \caption{Overview of an example AAC interface to create humorous comments (\emph{Keywords} interface). Screen 1: (a)~Joke Mode (b)~play text through TTS. Screen 2: (c)~keywords extracted from the conversation (d)~input field to type the starter of the joke (e)~dynamically updated suggestion (f)~get new suggestion}
    \label{fig-walkthrough}
\end{figure}

Before diving into specific UI design considerations, we demonstrate the general concept of one of our applications to get a sense of the UI flow in a brief overview outlined in Figure~\ref{fig-walkthrough}. For research purposes, we designed our interfaces to be AAC device-independent, running in the browser and basically consisting of a text field where users can type and access the Text-To-Speech (TTS) module. By keeping a simple interface and implementing it as a browser tool, we enable users to leverage their own input devices (ranging from eye tracking to virtual or physical keyboards, as we have seen in our qualitative interviews). All four interfaces share the same back-end, which captures the conversation context using the device's microphone and processes the content using the LLM. 

(a)~We extended this base app with a ``joke mode'' which triggers the different forms of interfaces we explore in this section to support users in composing humorous comments. (b) Users can press the button to output their humorous comments through text-to-speech. (c)~This example interface, later referred to as \emph{Keywords} interface, displays keywords based on the ongoing conversation, which is captured and transcribed using automatic speech recognition (ASR). (d)~Users either start typing their joke and/or select keywords to compose it. (e)~The interface displays auto-completed suggestions for jokes and updates these as users provide more context. (f)~ If users dislike a suggestion, they can hit the refresh button to get a new one. Once the user is satisfied with a suggestion, they select it and copy it into the input field. From there, they can either play the text immediately or make further customizations.

\subsection{Interface 1: AKA \emph{Context Bubble Selection}}
\begin{figure}[H]
  \centering
   \includegraphics[width=0.32\textwidth, alt={}]{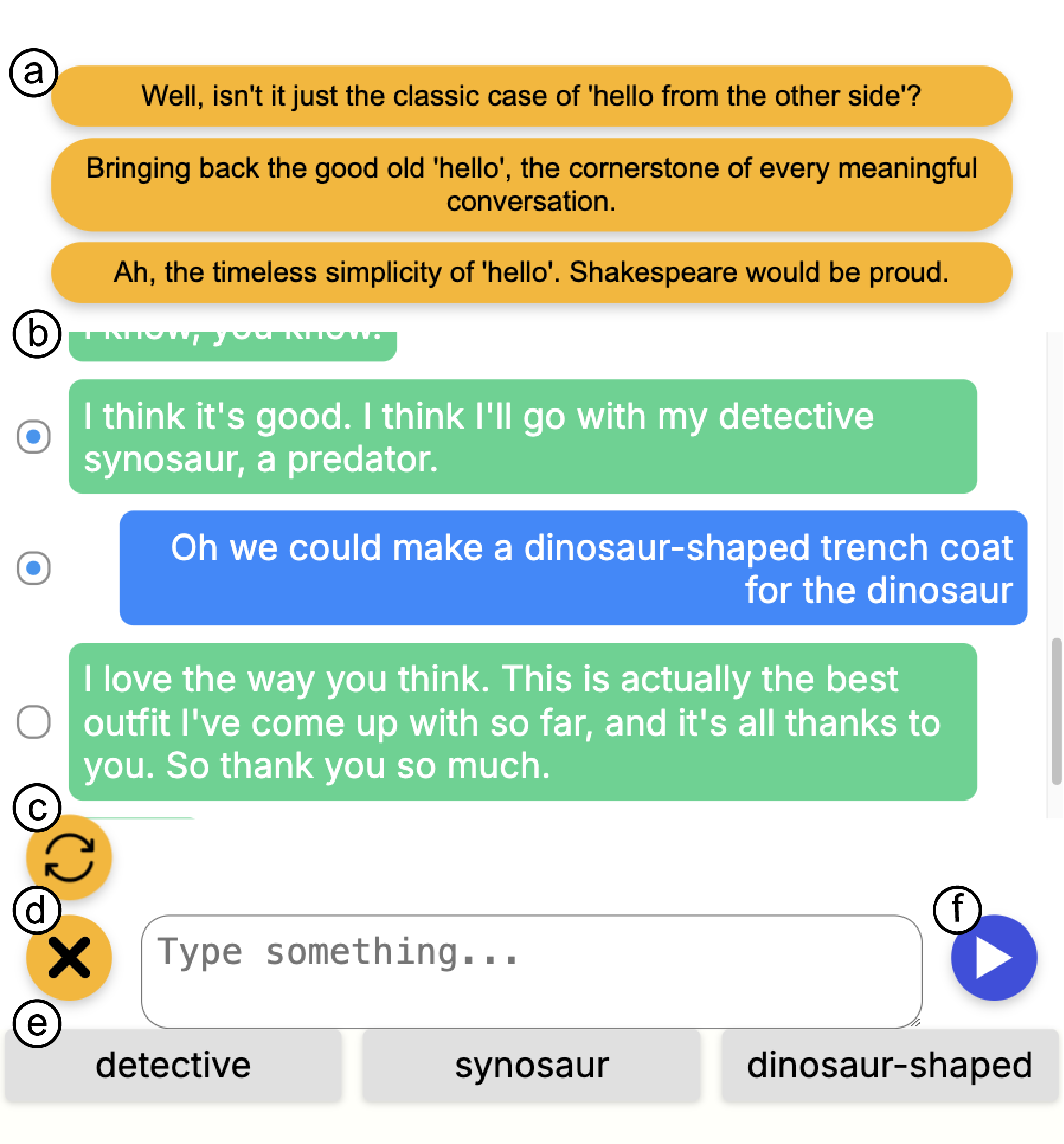}
  \caption{ Interface 1--\emph{Context Bubble Selection} features (a)~dynamically updated suggestions (b)~context selection from the transcript (c)~new suggestions (d)~exit joke mode (e)~keywords extracted from the selected bubbles (f)~output through TTS.}
  \label{fig-interface_1}
\end{figure}

Users compose humorous comments by selecting one or more bubbles from the conversation to set the context for their joke or comment, which results in suggestions from the LLM displayed in Figure~\ref{fig-interface_1}a. To further tune the jokes, the interface displays keywords extracted from those speech bubbles (Figure \ref{fig-interface_1}e).

Additionally, users can type directly into the text input field; any text they enter will be prioritized when generating suggestions. When users click on a suggestion, the text is transferred to the input field, allowing them to either edit it or play it using text-to-speech.

This interface enables users to interact directly with specific content of the ongoing conversation. The interface resembles familiar messaging applications, displaying the conversation transcript as speech bubbles.

\subsection{Interface 2: AKA \emph{Keywords}}
This interface more generically guides users in composing humorous comments by keyword interaction. The interface extracts six keywords from the last five sentences of the conversation. The interface generates one general suggestion, which users can refine by selecting one or more of the extracted keywords. If the user is not satisfied with the current suggestion, they can hit the refresh button to generate new suggestions.

\begin{figure}[H]
  \centering
  \includegraphics[width=0.32\textwidth, alt={}]{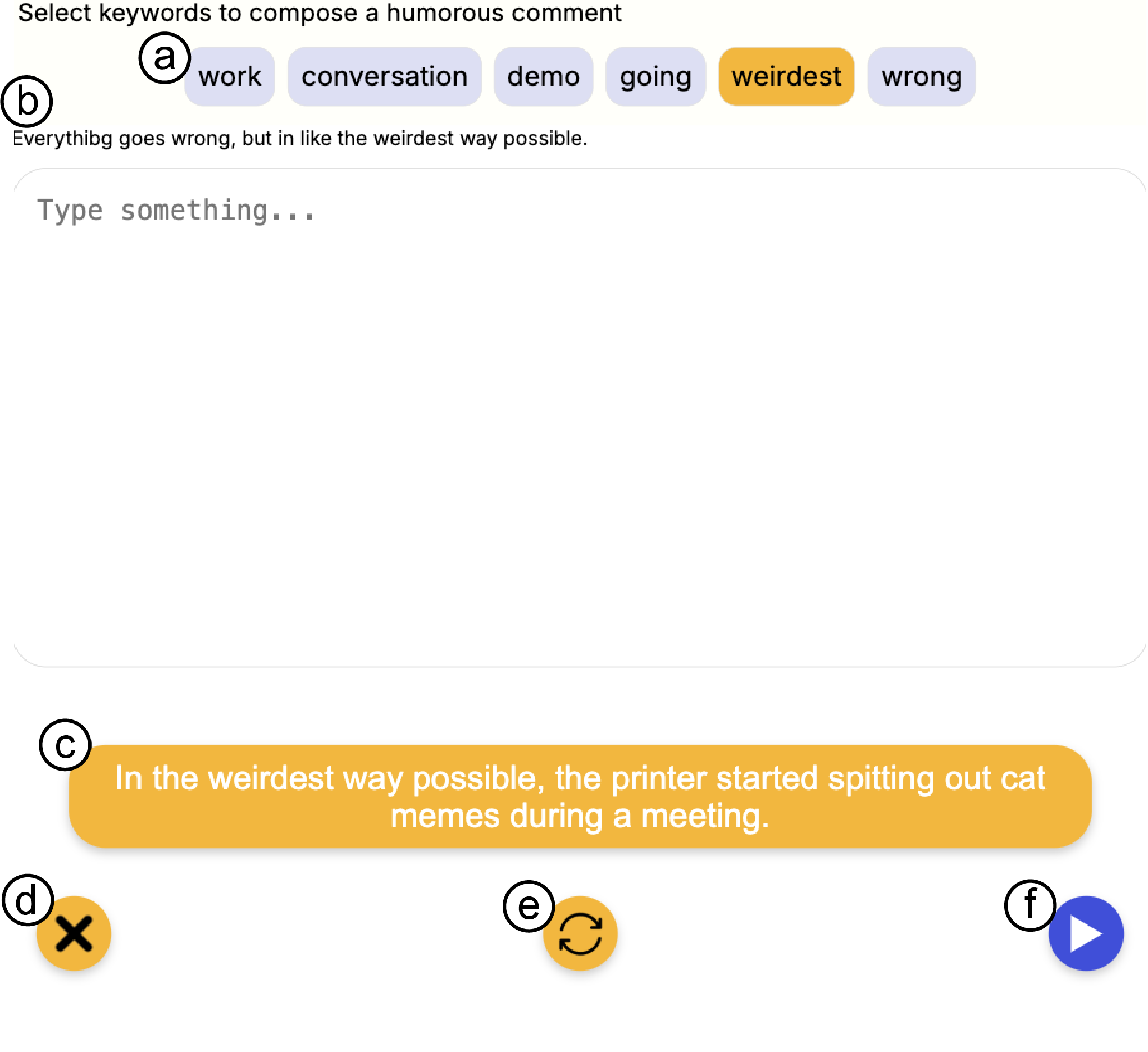}
  \caption{Interface 2 - \emph{Keywords}, features (a)~keywords extracted from the conversation (b)~transcript from the conversation (c)~dynamically updated suggestion (d)~exit joke mode (e)~get new suggestion (f)~play text through TTS.}
  \label{fig-interface_2}
\end{figure}
Like in \emph{Context Bubble Selection} interface, users can type directly into the text field, with any text they enter prioritized when generating suggestions. When the user selects a suggestion, the text is transferred to the input field, where they can edit it or play it TTS.

\subsection{Interface 3: AKA \emph{Wizard}}
This interface is inspired by the Witscript algorithm~\cite{toplyn2023witscriptgeneratingimprovisedjokes}, which prompts users to select from a set of keywords (based on the last 5 sentences of transcribed conversation). Subsequently, users select associations with those keywords. The interface responds by displaying three possible humorous comments based on the chosen keywords and associations. These suggestions are dynamically updated when the user selects keywords or associations.

This interface does not allow users to edit the suggestions or start typing instead. When a suggestion is clicked, it is immediately played through text-to-speech for efficient delivery.

\begin{figure}[H]
 \centering
  \includegraphics[width=0.32\textwidth, alt={}]{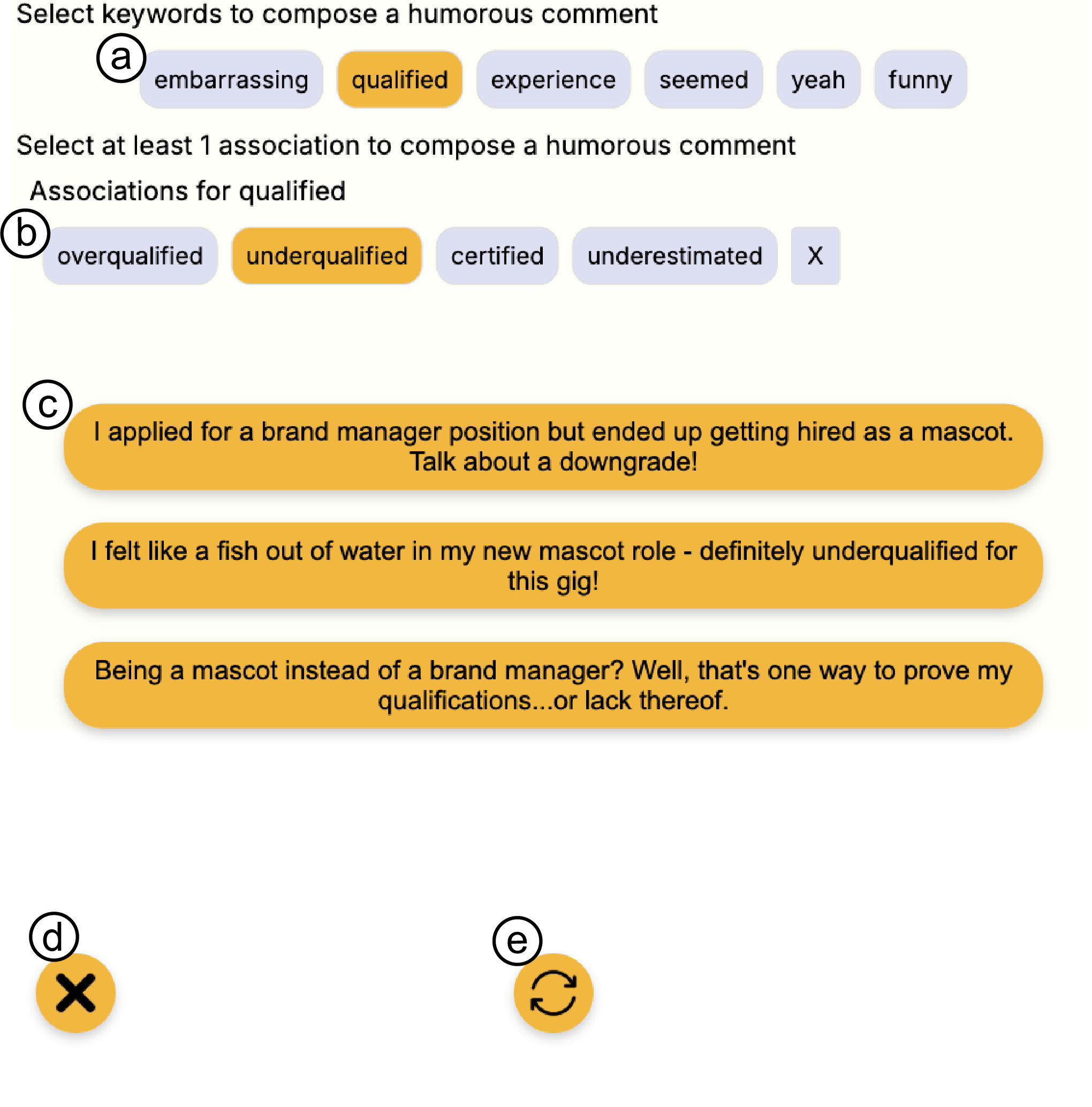}
  \caption{ Interface 3--\emph{Wizard} features (a)~keywords  from the conversation (b)~association from the keywords  (c)~suggestions (d)~exit joke mode  (e)~ refresh suggestions}
 \label{fig-interface_3}
\end{figure}

\subsection{Interface 4: AKA \emph{Full-auto}}

The Full Auto interface prioritizes efficiency over agency. By placing almost all decision-making in the hands of the LLM, this interface minimizes the user’s role, allowing for the quickest possible response in generating a humorous comment. Users can still refresh suggestions or make minor edits, but most of the interaction is automated. 

This interface places \textit{almost} everything in the hands of the LLM. The interface simply displays what it deems the ``best'' humorous comment based on the transcribed conversation.
\begin{figure}[H]
  \centering
  \includegraphics[width=0.32\textwidth, alt={}]{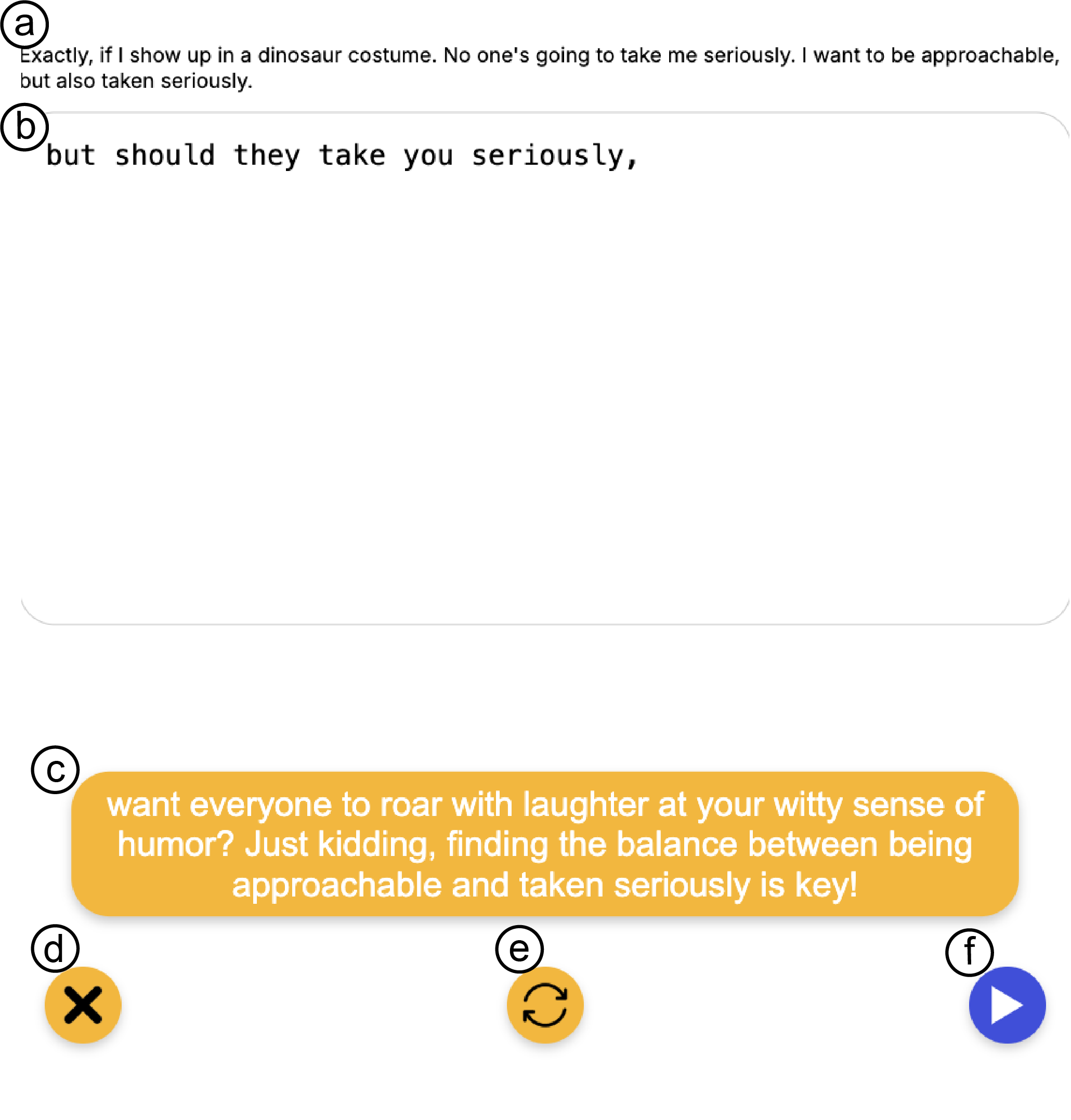}
  \caption{ Interface 4 - \emph{Full-auto}: features (a)~transcript from the conversation (b)~manual~input text (optional)   (c)~best suggestion based on the conversation (d)~exit joke mode  (e)~get new suggestion (f)~play text through TTS.}
  \label{fig-interface_4}
\end{figure}

\subsection{Design Components}

There are three dimensions of user interface components in the simple interface introduced above. We explore this by developing four applications that span the different dimensions. Table~\ref{tab:interfaces} shows how each interface represents a subset of the different UI components. The design of these components is detailed in the next sections.

\begin{table*}[t]
    \centering
    \begin{tabular}{l|cccc|ccc|ccc}
        & \multicolumn{4}{|c}{\textbf{tuning of suggestions}} & \multicolumn{3}{|c}{\textbf{suggestions}}  &
         \multicolumn{2}{|c}{\textbf{guidance}} \\
         Interface &context bubbles& keywords & associations & auto & single & multiple & editable & open & wiz & \\
\hline
         
         Context Bubble Selection& $\bullet$ & $\bullet$ &  &  &  & $\bullet$ & $\bullet$ & $\bullet$ & &\\
         Keywords&  & $\bullet$ &  &  &  $\bullet$&  & $\bullet$ & $\bullet$ & &\\
         Wizard&  &  $\bullet$& $\bullet$ &  &  & $\bullet$ & &  & $\bullet$ &\\
         Full-auto&  &  &  & $\bullet$ & $\bullet$ & & $\bullet$ &  & $\bullet$ &\\
    \end{tabular}
    \caption{The different interfaces and their UI components.}
     \label{tab:interfaces}
\end{table*}

Table \ref{tab:interfaces} summarizes the features of each interface \emph{Context Bubble Selection} interface lets users interact with the conversation context by selecting speech bubbles and keywords to tune the jokes. \emph{Wizard} is closer to \emph{Keywords} interface in that it also uses keywords as the base, but requires additional associations related to these keywords to fine-tune their meaning in the context of a joke. Finally, \emph{Full-auto} is the simplest interface, it just generates suggested jokes automatically. It uses the conversation history and gives the "best" humorous comment to respond.

\subsubsection{Tuning of Suggestions}

\begin{figure*}[h]
  \centering
  \includegraphics[width=\textwidth, alt={Tuning of Suggestions, Option one to the left (a) no options to tune, take it or leave it, Option two to the left(b) tuning through keyword selection, Option three to the right (c) tune by selecting context of the ongoing conversation and Option four to the right (d) selecting keywords to tune and further detailing them through associations}]{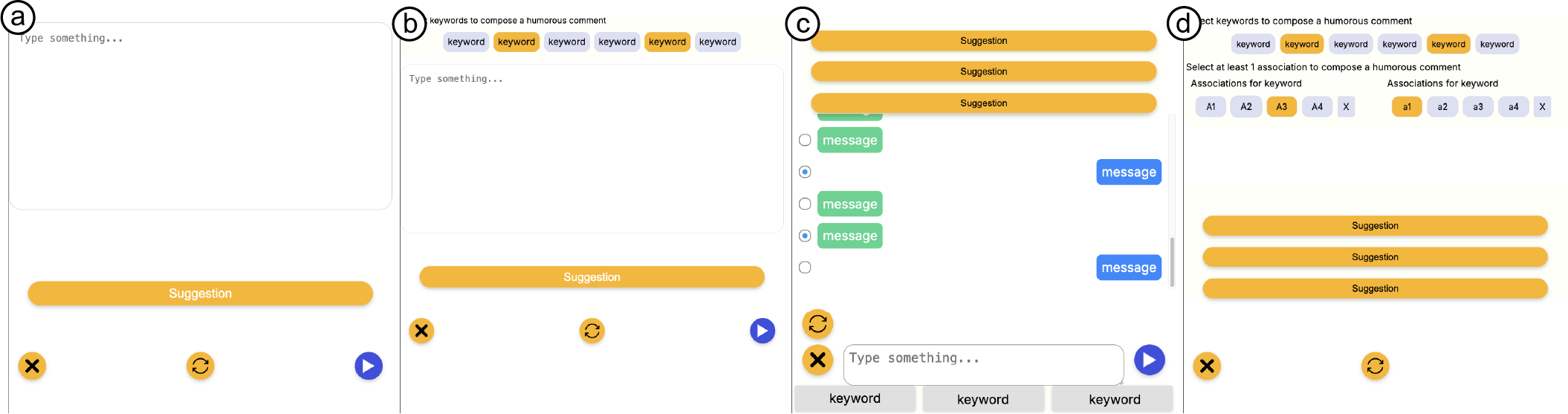}
  \caption{Tuning of suggestions, (a)~no options to tune, take it or leave it (\emph{Full-auto}), (b)~tuning through keyword selection (\emph{Keywords}) (c)~tune by selecting context of the ongoing conversation(\emph{Context Bubble Selection}) and (d)~selecting keywords to tune and further detailing them through associations (\emph{Wizard}).}
\label{fig-input-technique}
\end{figure*}

The input required by users to create humorous comments plays a large role in the efficiency of using auto-complete. We explore four different solutions. More user input will reduce the efficiency but give users an increased sense of control. Figure~\ref{fig-input-technique} shows the different techniques side-by-side. 
(a)~\emph{Full-Auto} is the simplest interface. It does not allow the user to tune the joke; it just creates a suggestion for a joke based on the current conversation context. 
(b)~\emph{Keywords} simplifies the interaction by distilling the conversation into essential terms, possibly leading to faster interjections and thereby enhancing efficiency while maintaining moderate agency through keyword choices. Both \emph{Context Bubble Selection} and Keywords interfaces allow for keyword selection. 
(c)~\emph{Context Bubble Selection} allows users to interact directly with specific parts of the conversation, offering high agency by providing detailed context for generating humorous comments. However, this comes at the cost of efficiency due to the increased interaction time required. Finally, (d)~\emph{Wizard}, inspired by \textit{Witscript}~\cite{toplyn2023witscriptgeneratingimprovisedjokes}, allows users to specify associations related to selected keywords to steer the joke better. The additional association helps disambiguate the way the keyword is to be used. 

Next to these different methods of tuning suggestions, all interfaces except \emph{Wizard} enable generating a joke by starting to type. This will specifically include the typed content as the start of each suggestion. By comparing these methods, we gain insight into how different levels of user control and interaction complexity impact the balance between expressiveness and speed in AAC interfaces.

\subsubsection{Presentation of Joke Suggestions}

We explore three ways to present the suggested jokes to users: Multiple Suggestions, Refining a Single Suggestion, and Looping Through Suggestions. Multiple Suggestions (\emph{Context Bubble Selection} and \emph{Wizard} Interfaces) enhance agency by offering a variety of options, increasing the chance that users will find a fitting comment. However, the time required to read several choices slows down decision-making, particularly in fast-paced conversations. Refining a Single Suggestion (\emph{Keywords} Interface) provides users with a quick starting point that can be refined. Before triggering TTS, the suggestion is displayed in the text field, allowing users to make small adjustments to tailor the output to their needs without sacrificing too much speed. Finally, all interfaces allow users to loop through suggestions if they are not satisfied with a specific joke. Combined, these strategies highlight how different design choices affect user experience in AAC interfaces, offering varied solutions.

\subsubsection{Open or Guided Interface}

\begin{figure}[h]
  \centering
  \includegraphics[width=0.5\textwidth, alt={UI guidance approaches: On the left a) Wizard-style fully guided interface, on the right b) Generating and Tuning Suggestions}]{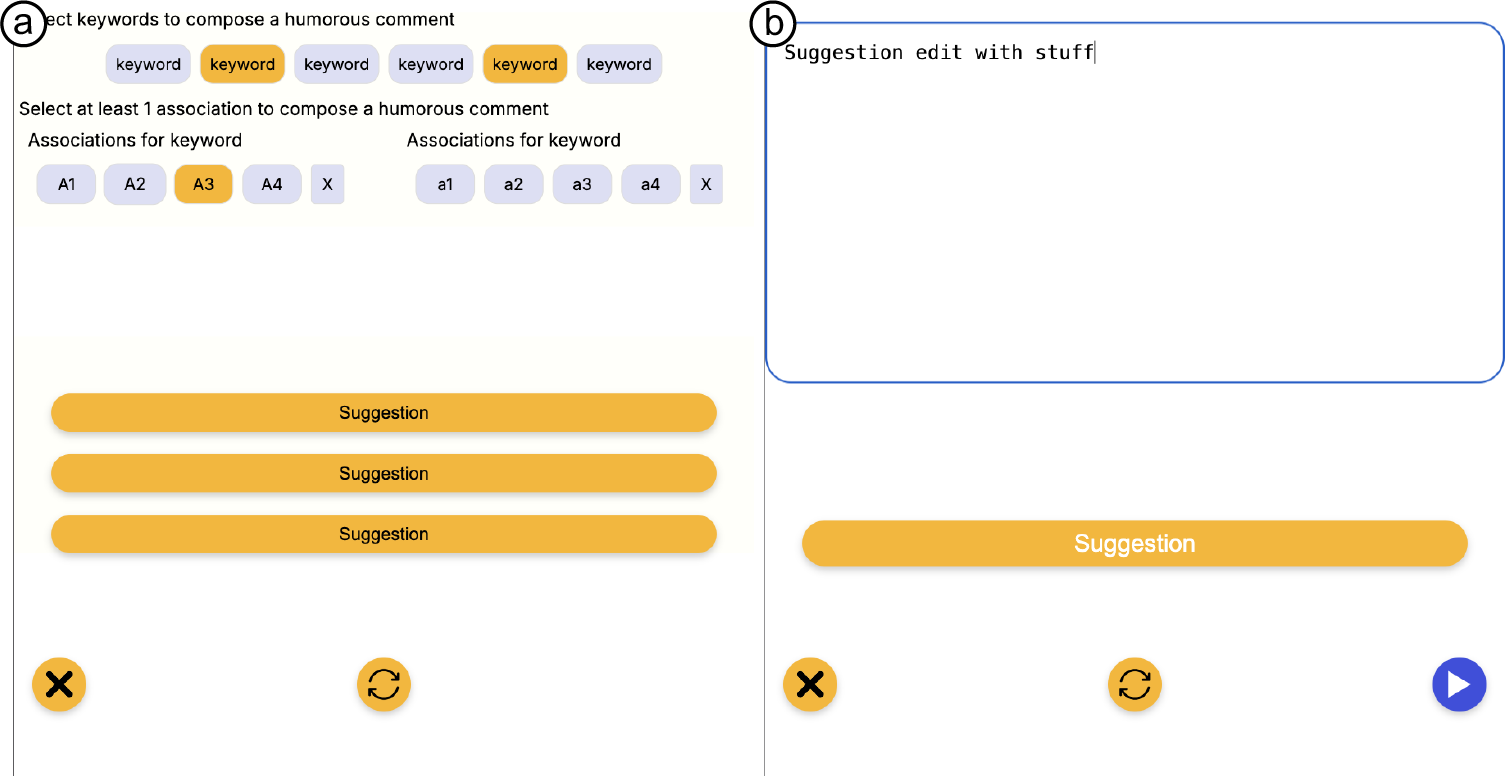}
  \caption{UI guidance approaches (a)~Guided: Wizard-style fully guided interface (b)~Open: Tune-styled which allow the tuning of generated suggestions}
   \label{fig-guidance style}
\end{figure}

Past work supports the idea that constrained choices can support ease of creative expression~\cite{tamburro2022comic}. We explore two guidance approaches: Wizard-Style guidance (Figure~\ref{fig-guidance style}a) and no/open guidance (Figure~\ref{fig-guidance style}b). The Wizard-Style guidance in (\emph{Wizard} interface) requires users to input keywords and associations to generate jokes. On the other hand, more open guidance (\emph{Context Bubble Selection} and \emph{Keywords} interfaces) provides users with quick, AI-generated suggestions that can be refined further through additional UI element selections if needed. While the guidance provides likely well-fitting jokes, the more open style refinement can be significantly faster at the cost of less structure in coming up with jokes.
\subsection {Implementation}
\begin{figure}[h]
  \centering
    \includegraphics[width=0.4\textwidth, alt={Shows system diagram containing our developed stack that used React and Tailwind for the front-end and Flas, NextJS, and RESTful API on the back-end, our stack connects to OpenAI GPT API and Amazon web services via HTTP request. On Amazon web service we use Amazon Polly for the text-to-speech and Amazon Transcribe for automatic speech recognition. On the other end our stack serves as a web application for the AAC users}]{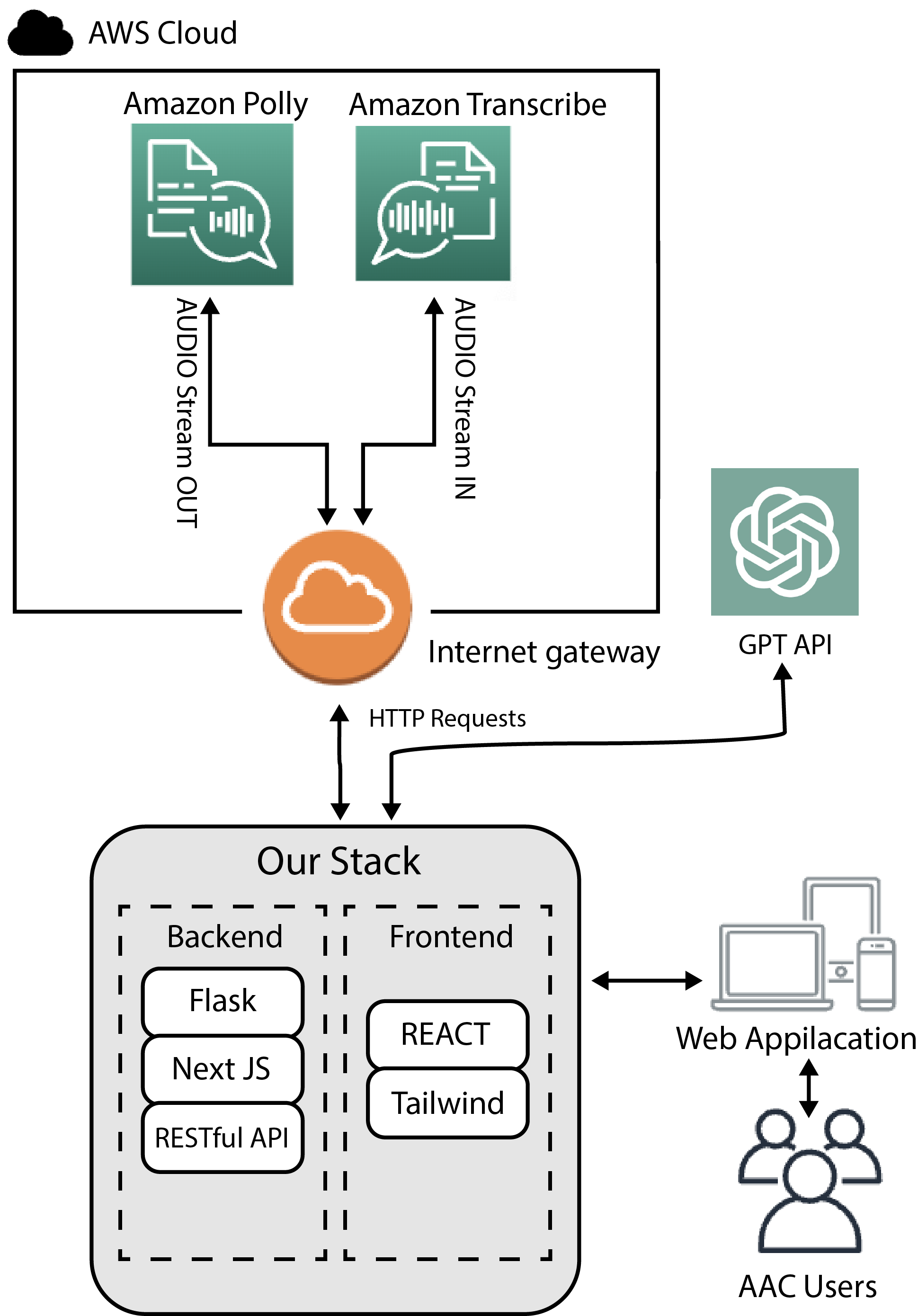}
  \caption{Overview of system diagram.}
  \label{fig-system}
\end{figure}

The interfaces described above run on the web, making them accessible from any device with an \textbf{internet} connection, which adds versatility and suitability for a wide range of AAC devices, including iPads, iPhones, and personal computers. This accessibility is especially beneficial for AAC users, as it provides flexibility to use the tool on whichever device is most convenient. The tech stack, illustrated in Figure~\ref{fig-system}, is made available to other researchers as a platform for exploring further interface design for AAC technology. At the core is a NEXTJS 14.2.5 framework running on Vercel, allowing for easy deployment across browsers. The front end uses React and Tailwind, while the back end consists of three core components: Amazon Transcribe for speech recognition, Amazon Polly for neural text-to-speech, and OpenAI GPT \cite{openai_gpt-4_2023} to process conversation, generate keywords, associations, and suggestions. During the user testing, we did not experience any noticeable lag with the GPT API as long as the device had a stable internet connection.

The system prompt is designed to generate contextually relevant humorous comments tailored to the user’s intent. By simulating the perspective of the user, the system prioritizes the context of selected conversational messages while incorporating user-provided keywords as secondary input. This approach ensures that the generated comments align closely with the ongoing interaction and fit into the ongoing conversation.

An example prompt from \emph{Context Bubble Selection}: "Imagine you are the user, generate 3 humorous comments start always with the user input and complete giving maximum priority to the context of the selected messages and second to the keywords. Give a short and concise sentence that the user could fit into the conversation"

%% file: src/04_evaluation.tex
\section{User Study: Understanding the practical implications of design choices}
We conducted a user study using the four interfaces to evaluate the impact on the user experience of the design trade-offs discussed in the previous chapter. The study and data collection protocols were reviewed and approved by our institution's IRB board and approved under protocol IRB\#0148755.

\subsection{Participants}
Five participants completed the user study. Due to personal circumstances, P6 and P7 could not participate. P1 and P3 participated in person, at their homes, which added an extra layer of contextual information about how they interacted with the researcher conducting the experiment, including non-verbal cues. For P3, whose fine motor skills were significantly limited and whose AAC device could not connect to the internet, a researcher acted as a proxy to facilitate interaction and brought their own device to run the prototypes. P3 generated messages on her AAC device and indicated their desired selections, which the researcher entered using a keyboard and mouse on the interface. All other participants interacted directly with the interfaces, using either their AAC device or personal device, with a range of input methods such as eye-gaze, direct touch, or keyboard and mouse.

\subsection{Procedure}
Each study session consisted of an interface demonstration followed by a conversational task with a researcher to use the corresponding interface. Participants engaged in four fictional conversations lasting approximately 6 to 12 minutes. The conversations included the participant and a researcher in a one-on-one format. The researcher role-played as a conversational partner while a second conversational partner (CP) was present in the room for protocol or system clarification if needed. However, the second CP did not actively participate in the conversation. P2-P5 experienced all four interfaces during the study, and P1 experienced all interfaces except the full-auto interface because they had to end the study early.

\subsubsection{Conversational Tasks}
The conversations were semi-structured, informal, and focused on generating humorous comments based on the ongoing dialogue. Study participants engaged in conversation in various ways: using either speech (via text-to-speech) or written text. We designed four conversational tasks to elicit humorous comments within the context of a real-time interaction:
\begin{itemize}
  \item \textbf{Task 1: Getting Lost in a Costume Store -} In this task, the researcher and the participant were trying to find the perfect outfit for an upcoming party. 
  \item \textbf{Task 2: Applying to the Wrong Job -} In this task, participants responded to a scenario where the researcher mistakenly applied for and received a job in a funny or unusual role. Participants were asked to offer advice on what the researcher should do next.
  \item \textbf{Task 3: Explaining a Funny Mishap at Work -} In this task, the researcher described a humorous mishap they experienced at the office to the participant who acted as a workmate. The task required participants to make light of the situation while the researcher explained what happened.
  \item \textbf{Task 4: Ordering Pizza -} In this task, participants engaged in a conversation where they were placing a pizza order with the researcher. The discussion involved humorous misunderstandings about toppings, delivery instructions, and the restaurant's unusual options.
\end{itemize}

\begin{figure*}[h]
 \centering
 \includegraphics[width=0.8\textwidth, alt={Contains results of 5 Likert scale questions. Question 1: How well did the humorous interventions match with what you tried to communicate? \emph{Context Bubble Selection}, \emph{Keywords}, \emph{Wizard}, and \emph{Full-auto} interfaces had 3 positive scores. Interface \emph{Keywords} and \emph{Wizard} had 2 negative scores. Interface \emph{Context Bubble Selection} and \emph{Full-auto} had 1 negative score. Question 2: How did you feel the timing of your humorous interventions were in the conversation? Interface \emph{Context Bubble Selection} had 4 positive scores and one neutral. Interfaces \emph{Keywords} and \emph{Full-auto} had 3 positive scores. Interface \emph{Wizard} had 2 positive scores and 2 negative scores. Question 3: Did you feel that the "joke mode" was effective in balancing expressivity and timing? Interface \emph{Context Bubble Selection} had 2 negative scores and two positive scores. Interface \emph{Keywords} had 3 negative scores, 1 neutral, and 1 positive. The \emph{Wizard} interface had 2 negative scores, 1 neutral, and 1 positive. The \emph{Full-auto} interface had 1 negative score, 1 neutral, and 2 positive. Question 4: Did you find the humor generation features helpful? \emph{Context Bubble Selection} interface had one negative score, one neutral, and 3 positive. \emph{Keywords} interface had 2 negative scores, one neutral and 2 positive. The \emph{Wizard} interface had 2 negative scores and 3 positive scores. The \emph{Full-auto} interface had 1 neutral score and 3 positive. Question 5: How effective was this interface in helping you create humorous comments? \emph{Context Bubble Selection} interface had 1 negative score and 4 neutral. \emph{Keywords} interface had 1 negative score, 2 neutral, and 2 positive. The \emph{Wizard} interface had 2 negative scores and 3 positive. The \emph{Full-auto} interface had 2 neutral scores, and 2 positive}]{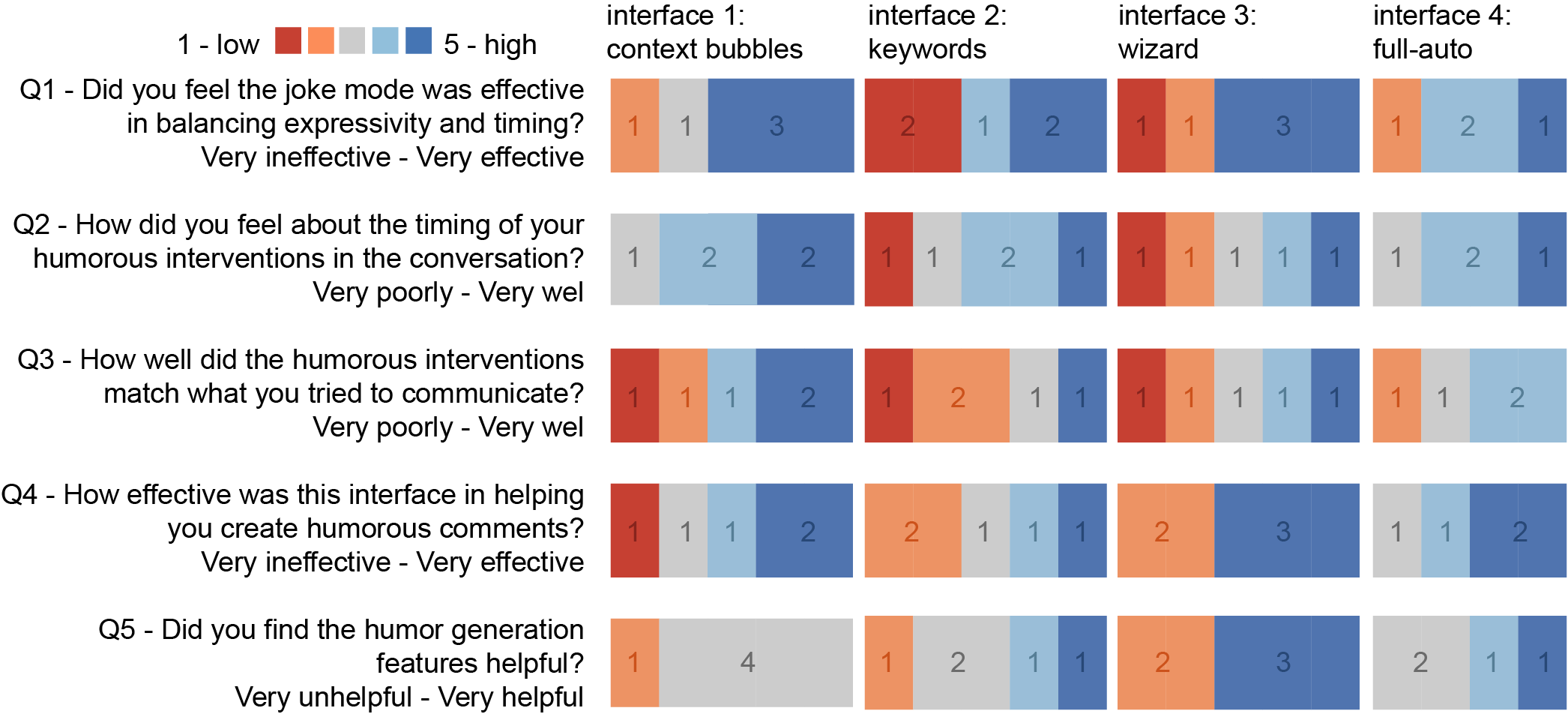}
 \caption{Summary of Likert scale questionnaire results.}
  \label{fig-survey-results}
\end{figure*}

After completing each task, participants filled out a Likert scale questionnaire to provide feedback. Finally, after the four conversational tasks, participants filled out a survey where they could elaborate on what they liked or disliked about the interfaces. Survey questions included open questions like: What did you like the most about the interfaces? What did you find most challenging about using the interface?

\subsection{Data Collection and Analysis}
Video and audio were recorded for analysis during each session. We recorded the participant's screen during the tasks to capture the interaction with each prototype UI. 
 
\subsubsection{Interaction Coding}
We analyzed the video recordings using detailed interaction coding, focusing on participants' interactions with the interfaces, including actions such as entering joke mode or engaging with specific features. Interactions were categorized by feature type and visualized using distinct colors and shapes (Figure \ref{fig-interaction-coding}). \edits{Following an inductive approach to video analysis~\cite{ibrahim2018design}, videos were repeatedly reviewed and indexed to identify key segments of AAC usage.} \edits{Following established interaction analysis methods~\cite{curtis2023watch}, we coded 327 instances. This systematic approach revealed emerging patterns in user interactions, enabling us to extract actionable conclusions to inform interface design.}

\subsubsection{Likert-scales and survey analysis}
We collected Likert scale questionnaires to assess the usefulness of each feature. We captured participants' thoughts on how well the tool performed in generating timely humorous comments (e.g., Did you find the humor generation features helpful? and How well did the humorous interventions match what you tried to communicate?). In Figure \ref{fig-survey-results}, we present the results of the Likert scale questionnaires.

\subsection{Results}

We present insights into participants' experiences with various interfaces designed to support AAC users to create timely humorous comments. We explored four interfaces proposed in our design space: (1) Context Bubble Selection, (2) Keyword Selection, (3) Wizard-style Guidance, and (4) Full-auto Suggestions.

In section 7.4.1, we present participants' user ratings (See Figure \ref{fig-survey-results}) and describe their overall experience with each interface. 
In section 7.4.2, we dive deeper into participants' patterns of interaction with each interface (See Figure \ref{fig-interaction-coding}) and describe how well each proposed feature supported AAC users, creating timely humorous comments.
In sections 7.4.3 and 7.4.4, we highlight participants' preferences for the interfaces and features and recommendations participants shared to improve future tools to support the creation of timely humorous comments with AAC.

\subsubsection{User Experience}
In this section, we present results of how participants rated each interface in terms of (Q1) balancing expressiveness and timing, (Q2) timeliness, (Q3) expressiveness to match their intention, (Q4) effectiveness, and (Q5) helpfulness (all relating to generating timely humorous comments). Surprisingly, some participants favored the Full-auto interface over other interfaces despite taking away their agency to generate jokes. Other interesting findings included: (1) Most participants found the Context Bubble Selection Interface the best in terms of intention expressiveness (matching their intention), as well as timing, striking the middle ground in this trade-off; despite this, (2) Most participants scored Context Bubble Selection neutral or somewhat negatively in terms of helpfulness to generate jokes. We provide more details in the following subsections.

\edits{\textbf{Q1: Balancing Expressiveness And Timing}}

 The \emph{Context Bubble Selection} interface ranked first in balancing expressiveness and timing (Q1), though not by a significant margin. All interfaces received three positive reactions, but the strongest positive reactions were for \emph{Context Bubble Selection,} with three participants giving the highest score (5) and one participant giving a neutral score. Participants appreciated how \emph{Context Bubble Selection} interface struck a reasonable balance between providing features that let them refine the joke to their intended message and the time it took to generate it. \emph{Keywords}, \emph{Wizard}, and \emph{Full-auto} interfaces all had an acceptable balance, though the \emph{Wizard} interface came slightly ahead over the \emph{Keywords} interface with two participants providing slightly better individual scores ( three 5s vs two 5s and one 4, and one 1 and one 2 vs two 2s). Participants weighted intention as more important than timing (as seen in our interview results!) to decide which interface had a better balance between intention and timing. While participants had to go through many steps to evoke their intention when using the \emph{Wizard} interface, the benefits of being able to refine their intention outweighed the sacrifice in timing when considering how well they are balanced. When comparing \emph{Keywords} and \emph{Full-auto} interfaces, there is no clear difference, though participants had stronger opinions for \emph{Keywords} interface ( two 1s and two 5s vs zero 1s and one 5). Again, we believe this was because the mismatch in intention when using \emph{Keywords} interface evoked stronger negative feelings overall.

\edits{\textbf{Q2 and Q3: Timing and Matching User Intention}}

\begin{figure*}[h]
 \centering
 \includegraphics[width=\textwidth, alt={Video coding of interactions: Sequences of interactions that are relevant to humorous comments compositions features by participant. Participants' data is grouped by Interface and then sorted by participant ID. Note that when the user generated comments without joke mode enabled was not in the coding for brevity. *P5 did not interact at all with I3 and **P1 was not able to complete the study for I4.}]{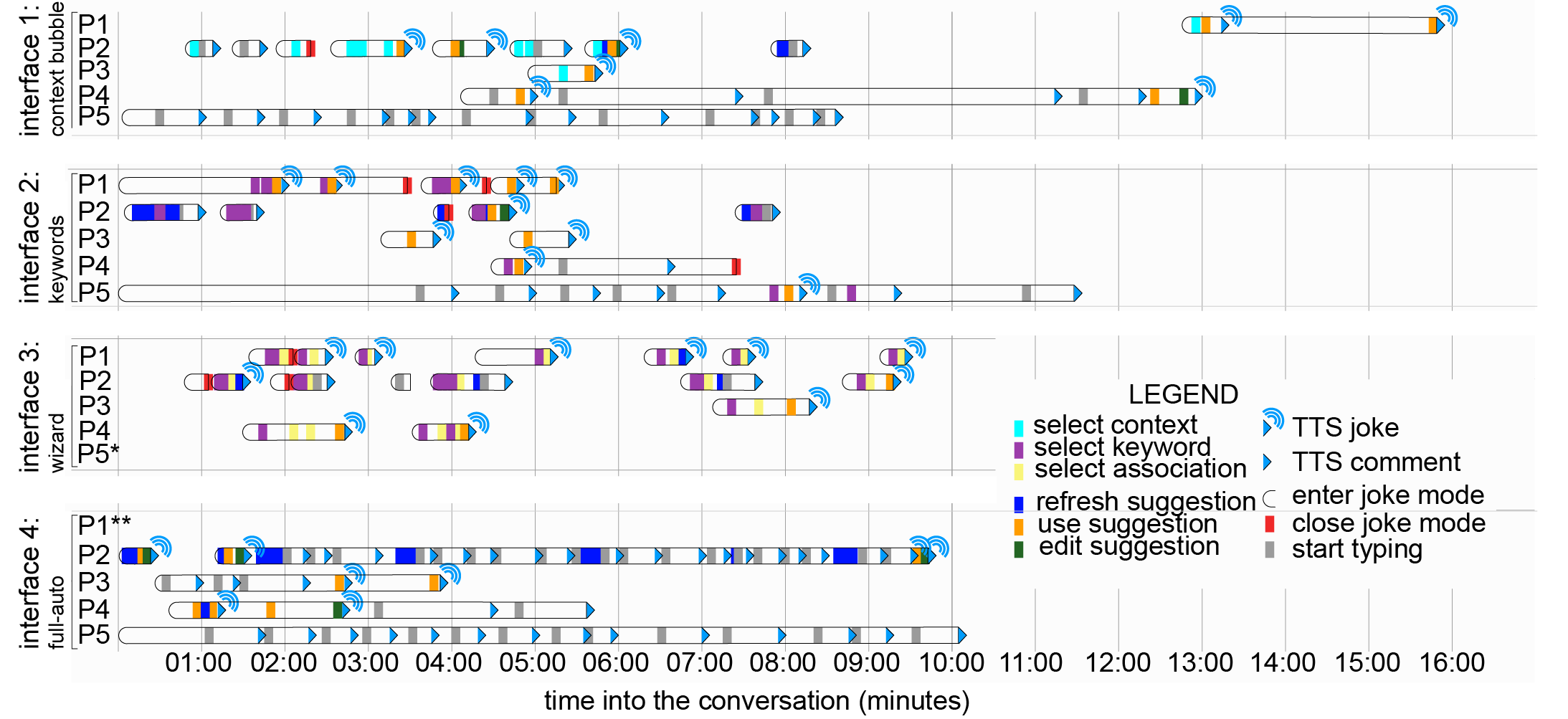}
 \caption{Video coding of interactions: Sequences of interactions that are relevant to humorous comments compositions features by the participant. Participants' data is grouped by Interface and then sorted by participant ID. Note that when the user-generated comments without joke mode enabled were not in the
coding for brevity. *P5 did not interact at all with Wizard interface and **P1 was not able to complete the study for Full-Auto interface.}
 \Description{}
 \label{fig-interaction-coding}
\end{figure*}

The \emph{Context Bubble Selection} interface performed best in delivering timely humorous comments (Q2). Four participants rated it positively, while one gave it a neutral score. Meanwhile, the \emph{Wizard} interface had the strongest negative reaction: Two participants scored it negatively, and one scored it neutral. Using the \emph{Wizard} interface, users had to do more clicks than with other interfaces to generate jokes. For both the \emph{Keywords} and the \emph{Full-auto} interfaces, participants felt roughly similar (slightly worse for \emph{Keywords}): Both interfaces received three positive reactions and one neutral reaction. This is to be expected since both interfaces require users to click a similar amount of times to generate a joke. For example, with the \emph{Keywords} interface, the user might select two keywords and then send a joke, while with the \emph{Full-auto} interface, the user might have to refresh the joke generation several times before sending it (see Figure~\ref{fig-interaction-coding}). Surprisingly, \emph{Context Bubble Selection} interface came on top. We observed that participants, even if they had to click more times than in \emph{Keywords} and \emph{Full-auto} interfaces, could immediately select the portions of the conversation they were interested in to make a joke, making it faster than waiting for the right keywords or joke to appear. P4 explained why the keywords and full-auto interfaces were less likely to match their intention: \emph{"I think because I'm a person who reacts to what is said instantly, it's difficult to wait for the AI to come up with humorous suggestions. Especially if it doesn't match what I was instantly thinking"}. P2 added, "Refreshing [asking for a new suggestion] sometimes made it better,r but reloading and opening the joke interface were also actions that took time." 

As for matching the user's intention (Q3), the \emph{Context Bubble Selection} interface came ahead of the rest by a considerable margin. It was the only interface where three participants scored positively. The context bubble selection feature's ability to select snippets of the conversation and then keywords derived from those snippets strongly enhanced users' abilities to evoke their intention through the interface. The \emph{Wizard} ranked second, outperforming the \emph{Keywords} interface because it allowed users to refine their intention by using the suggested associations. Interestingly, at the lower end, the \emph{Full-auto} interface had marginally better scores than \emph{Keywords} interface even though users had less direct control of the joke output (they could only hit a refresh button that provided a joke based on the past 5 sentences). Three participants had negative reactions to \emph{Keywords} interface, while only one participant had a somewhat negative reaction to the \emph{Full-auto} interface. We observed that during their use of \emph{Keywords} interface, participants were frustrated because they did not have control over which keywords popped up, and sometimes the keywords they wanted to use did not appear immediately after the other person spoke. P2 highlighted this frustration: \emph{"It gave a lot of jokes that used the keywords but didn't actually fit the conversation"}. We observed that this mismatch in intention evoked stronger negative feelings for \emph{Keywords} interface.

\edits{\textbf{Q4 and Q5: Effectiveness and Helpfulness of Prototypes Generating Humorous Comments } }

The \emph{Context Bubble Selection} interface outperformed the \emph{Wizard} interface in Q3 (matching intention), but it was rated the least helpful (Q5) for generating humorous comments. While users could intentionally select concrete portions out of the conversation (thus, the best in Q3), some users found it confusing and struggled to select keywords from the selected conversation bubbles. This made it difficult for the LLM to generate appropriate jokes since users did not communicate to the LLM what it should focus on. For example, P4 and P5 had difficulty scrolling through the conversation history due to their input methods. Thus, while providing added agency, four participants scored \emph{Context Bubble Selection} interface as neutral and one as not helpful to generate jokes. 

Three participants scored the \emph{Wizard} interface as very helpful (Q5) and very effective (Q4) in generating timely humorous comments. Since participants managed to select contextually relevant keywords, and then associations to better match their intention, they had more control over what the LLM  used to generate jokes. This is reflected in its previous score on intention (Q3), where the \emph{Wizard} interface scored second best. Two participants scored this interface as not helpful and somewhat ineffective. This was reflected in the interaction pattern of these participants. For example, P3 engaged with the feature only once during their task interaction (See Figure~\ref{fig-interaction-coding}, Wizard-P3).

Interestingly, the \emph{Full-auto} and \emph{Keywords} interfaces both received only positive or neutral reactions: One very helpful, one helpful, and two neutral scores. As previously argued, both of these interfaces required a similar (low) amount of clicks to generate a joke. This resulted in the \emph{Full-auto} interface being perceived as effective (Q4); receiving two very effective, one effective and one neutral score. As P4 put it: \emph{"The Full-auto was the best of the interfaces so far due to simplicity"}. Despite taking some control from them in generating the jokes, some participants found interfaces that required fewer clicks to be helpful and effective in generating humorous comments.

\subsubsection{User Interactions.}
Here we present how participants interacted with each prototype. Some participants used the Context Bubble Selection Interface as inspiration to write their own humorous comments. However, others faced difficulty interacting with it due to their input methods (e.g., eye tracking). For the Keywords interface, all participants interacted with the keywords feature to compose humorous comments. One participant reported feeling they were not in control when using the interface. For the Wizard interface, while participants selected many associations, the increased number of interactions did not translate into a higher number of delivered jokes. One participant noted that there were "too many steps" to generate a suggestion. For the Full-Auto interface, participants appreciated its simplicity and ability to provide suggestions. In the following subsections, We provide more details about how participants' interactions differed with each interface.

\edits{\textbf{Interactions with \emph{Context Bubble Selection Interface}}:}
\edits{An exemplary interaction with the \emph{context bubble selection} interface was P2 talking about finding a costume for a party. Before engaging, they scrolled through the transcribed history and selected an earlier comment where they confused a trench coat for a dinosaur costume. The interface suggested (among two other comments): "Well, going from a trench coat to a dinosaur costume is quite the wardrobe malfunction!" which P2 liked but added "... or is it?" adding a personal sarcastic touch to customize it to their style (they used more sarcasm throughout).}

Participants' interaction with the conversation bubbles varied significantly (Figure \ref{fig-interaction-coding}, Interface 1). For example, P1 and P3 only selected the conversation context bubbles one time each to try it. P1 then used one additional joke suggestion without selecting the conversational context bubbles again. P3 liked that she could see the transcript of the conversation while listening to it, noting:: \emph{"It helped me think of jokes."}

P4 and P5 had a hard time scrolling through the conversation history due to their input methods (joystick mouse and eye gaze respectively) which interfered with their ability to see the whole screen. They did not use the conversation context selection feature. On the other hand, P2 selected conversation context bubbles 10 times during the task. Interestingly, P2 often used the interface to view humorous comment suggestions but typed their own version of the joke instead, as if they used suggestions only as inspiration support. All participants except P5 used the suggestions to share humorous comments.

\edits{\textbf{Interactions with \emph{Keywords} Interface}:}
\edits{P1 took advantage of the keyword selection in a conversation about \textit{a working mishap where the researcher tripped over a plant, and their slides were not loading}. By selecting three keywords, "slides", "work", and "situation", they interjected: "If I were you, I'd just embrace the dirt and slide into the situation head first - it can't be any worse than trying to get those slides to work!" emphasizing the ridiculousness of the situation and accompanied the comment with a cheerful laugh. This showcases the potential of personalized responses to amplify the user's expressive capabilities.}

All users were able to compose a humorous comment with this interface. P2 interacted with the interface most frequently, selecting 15 keywords throughout the conversation but, similar to the \emph{Context Bubble Selection} interface, they used the suggestions as a source of inspiration. After selecting a couple of keywords, they would turn off joke mode and then type their own version of the jokes. They only picked a suggestion once, and they also edited it before playing it.

We observed a recurring pattern among P3, P4, and P5: choosing only one keyword and picking the first suggestion. P1 usually selected 3 keywords to guide the AI to the joke she wanted. Interestingly, at the end of the conversation, P1 fired off two suggestions in a row without selecting any keywords. This unique behavior was not observed in any other interface interaction. Post-task P1 commented:\emph{"It was easy to use, it made sense but sometimes I felt like I wasn't in control"}. When asked how it could be improved, they added that \emph{"typing the keywords"} could help them feel in control. P3 added that she liked the \emph{"very basic"} interface. 

\edits{\textbf{Interactions with \emph{Wizard} Interface}:}
\edits{An example of a typical interaction with the \emph{Wizard} interface occurred during a conversation about costumes. The participants were discussing the shirt they were going to choose for the party, but the researcher unexpectedly selected a full-body dinosaur costume instead. P1 selected the keyword "shirt" and the association "size", but the suggestions were too focused on the dinosaur costume (since it was the last bit of conversation), so she hit the refresh button and chose the suggestion "Well, that shirt must have been T-rex size!" which combined all the elements she wanted, enabling her to maintain agency while achieving timely and contextually appropriate humor.}

This interface showed an increased number of interactions from P1, P2, and P4 compared to the rest of the interfaces, likely caused by the larger number of steps required to use it. P4 explored the generation of suggestions by selecting different associations for the same keyword and later by selecting two keywords and then 1 association. Instead of picking the second association, he selected 1 more keyword (previous step) and then selected the second association before finally choosing the suggestion. Showing a playful exploration of their control over the AI suggestions.

In a more standard way, P1 was able to share 6 jokes following the guidance of the interface by first selecting keywords, followed by associations, and finally picking a suggestion. Notably, P1 used the refresh option to obtain more suggestions, rather than choosing the first one as in previous interfaces.

P3 exhibited a very limited interaction with this interface, engaging joke mode only once. This may be because P3 relied on a researcher as a proxy to interact with on-screen elements, and this guided interface required many clicking steps before a user could get to joke suggestions. Similarly, P5 did not engage joke mode at all for this interface. We hypothesize that this guided interface might not be preferred for gaze input, but without further testing, this might be an abrupt assumption.

P2 interacted with this interface much more than all the others, engaging joke mode 7 times, selecting 11 keywords, and choosing 12 associations in total over the conversation. However, most of their interactions concluded with them closing joke mode and typing their own jokes after going through the whole guided process.
After the task, P4 expressed his discomfort of having \emph{"too many steps"} before he could get the suggestion. Furthermore, P1 shared \emph{"keywords [and associations] are fun to pick, but sometimes it's wasn't easy to choose [between so many options]."} Lastly, P3 expressed her excitement about this interface, which worked similarly to how her AAC device works, using associations on a symbol-based keyboard. 

\edits{\textbf{Interactions with \emph{Full-auto} Interface}:}


To illustrate the interaction with the full auto interface, we highlight an example during a conversation about pizza and pizza toppings. When asked for her opinion on pineapple on pizza, P3 opted to let the AI generate a response:``I think pineapple on pizza is like a funky dance party in your mouth - unexpected, but strangely delicious!". We were surprised by this choice since it was a "personal preference" question but she chose to answer with a quick joke. This suggests that users may sometimes prioritize delivering a timely and entertaining remark over directly addressing the conversational prompt.

P2 and P4 took advantage of the refresh button to quickly iterate through several suggestions before picking one, and they both opted to slightly tweak the suggestion with the manual input to make it more personal. For example, P2 hit the refresh button at least 24 times during the session but only used the suggestion three times with edits. For all other instances, P2 refreshed the suggestions and then proceeded to directly type their own humorous comment. 

In contrast, P3 never used the refresh button. From the two occasions when she picked a suggestion, she chose the first option without making any edits to be able to deliver the joke as quickly as possible. P5 completely ignored the suggestions and typed 100\% of his comments. We believe that this was due to the fact that when he opened his eye gaze keyboard to speak it overlaid over the part of the screen where the suggestions were. Sadly due to time constraints, P1 wasn't able to try this interface. 

P3 and P4 expressed that they liked the \emph{Full-auto} interface the most due to its simplicity. P3 shared: \emph{"It helped me think of jokes"}, adding that they liked that it gave suggestions without requiring her to click anything. When asked how well the AI matched her intentions, she replied: \emph{"It matched my intentions and also influenced it".} Similarly, P4 described it as the \emph{"best of the interfaces so far due to simplicity"} but also noted that: \emph{"As the other interfaces it takes time getting used to using the app generating thoughts and the delay of using the mouse delays flow [of the interaction]"}

\subsubsection{Overall Feature Preference}
P2 expressed that they liked the ability to edit the suggestion after picking it \emph{"It was a useful feature when it existed."} They also liked seeing both the jokes and the text input at the same time. On that note, P4 expressed that it was hard to use the applications on a separate screen from his AAC device, saying: \emph{"Seeing it on a device I can touch and be in line with my sight"} since he was using the computer for the Zoom and interfaces and his AAC device (touchscreen) as the virtual keyboard on the computer. Despite this, he liked the "Ease of learning" of the different interfaces. P3 expressed that they "really liked the first interface [\emph{Context Bubbles Selection}]", and "found the text-to-speech Setup most effective". P5 remarked: "It was "new" to me. Everything takes a moment to learn at first". P3 expressed a similar feeling, stating: "I just found learning several different interfaces hard."

 \subsubsection{Recommendations for improvements}
Participants expressed enthusiasm for the different range of options that our interfaces presented, as well as recommendations for access considerations and improvements. Both P2 and P4 expressed the need for "Full keyboard navigability" to which P4 expanded:``Keyboard shortcuts and things without a mouse in mind. Interfaces might be better on touchscreen for people using direct select methods, for scanners and eye gaze may be troublesome". This highlights an important consideration for future improvements. P3 suggested the inclusion of "an adult version and a child version for everyone", implying how she might want to use different types of humor in different situations. P5 added \emph{"A sarcastic voice would be nice"}. In general, all the participants expressed their excitement about the opportunity to test different interfaces. P3 encapsulated that feeling by closing the session with \emph{"I'm just excited to see where this project goes, and whether or not my future device will have it."}

%% file: src/05_futurework.tex
\section{Discussion}

Our work presents an early exploration of the design space of interfaces for AI-powered apps to support AAC users in making timely humorous comments. Our interview findings revealed the significance of humor in enabling AAC users to form social connections. We identified several barriers in existing AAC software and hardware that hinder AAC users' ability to generate humorous comments, such as timing, intonation, and technical limitations. Our user study findings revealed AAC users' preferences for generating timely humorous comments 'ratings revealed unexpected trade-offs, such as participants enjoying using the \emph{Full-auto} interface, even though it reduced their sense of agency. We also present user patterns of interaction, which demonstrate how each participant uniquely interacted with each interface. In the following sections, we reflect on the trade-off (agency vs. efficiency) and AI influencing our intentions. We provide design recommendations for developing future adaptive AAC technology.

\subsection{Balancing Agency and Efficiency}

Our study results highlight a trade-off between agency and efficiency in making humorous comments during conversations. 
We explain how these findings contrast with prior work and suggest that future designers create systems that can adapt to users' unique abilities. 
Further investigation is needed for other time-constrained tasks, such as back-channeling in ongoing conversations.

Participants rated the \emph{Full-auto} interface more positively overall, demonstrating that, when the timing is critical, there is a willingness to ``trade-in'' agency to deliver humorous comments faster.  We hypothesize that this preference stems from the significant speed-up offered by the Full-auto interface, coupled with the importance of timing for humorous comments, as identified in our interviews. Together, these factors appear to outweigh the perceived loss of agency in this context.

This finding contrasts with prior research, where AAC users hesitated to delegate decision-making to AI \cite{valencia_conversational_2020, valencia_less_2023}. Since users did not have time pressure to program their AAC \cite{valencia_conversational_2020, valencia_less_2023}, users valued maintaining control over their communication, prioritizing expressivity and personal authenticity over speed~\cite{kane_at_2017}. In our interviews, we observed a similar tension between expressivity and timing, but our participants leaned more toward timing when delivering humorous comments. This highlights the dynamic nature of this trade-off in the context of humor: while users may favor agency in slower, more deliberate interactions, they are more inclined to embrace automation in fast-paced, high-pressure contexts.

Following the principles proposed by the Ability-based Design philosophy~\cite{wobbrock2011ability}, future AI-powered AAC devices could be designed to dynamically adapt to meet users' unique needs. Adaptive systems could change their UI and functionalities by observing user behavior~\cite{gobert2019sam}, and by following instructions provided by users like their preferred interaction style and humor style. This adaptability could be especially helpful for AAC users, whose abilities may change significantly over time. With the rise of runtime AI-generated interfaces~\cite{wu2024framekit}, we envision future AAC systems capable of striking the right balance of agency and efficiency to meet each AAC user's unique needs and moment-to-moment conversational demands.

\subsection{AI Influencing Our Intentions}
AI systems offer significant efficiency gains and assistance, but they can also shape user behavior in subtle ways~\cite{williams2024bias}. AAC users are particularly vulnerable to this AI phenomenon. For example, during our user study, P3 remarked on the AI's ability to match their intentions while simultaneously influencing them during the interviews, P4 reflected on the relationship between an AAC user and their device in shaping their ``inner voice''. This highlights the vulnerability of AAC users— if they begin adopting the vocabulary and style of an LLM, the boundary between their self-perception and the device could blur further, potentially leading to a loss of self-expression and authenticity.

Self-expression and authenticity are key goals for AAC users~\cite{kane_at_2017}. This raises a challenging question: how can we ensure that users' self-expression remains authentic and aligned with their intentions when using AI-powered AAC? Striking a balance between leveraging AI's capabilities and preserving genuine self-expression remains a complex challenge for designers of AAC technologies. One possible approach could involve the development of highly personalized AI models trained on individual speech patterns and preferences. However, this solution may come with its risks: It could further entrench the influence of AI on personal expression, subtly shifting intentions over time. 

AI offers the potential to enhance timing and efficiency but risks influencing users’ intentions and diminishing their authenticity. To address this, designing systems that provide support without overriding individuality will be critical to maintaining the agency and authenticity of AAC users in their interactions. systems must support individuality by prioritizing personalization, expressivity, and adaptability. This includes going beyond generic AI outputs to handle unpredictable, context-specific interactions, like humor, without overriding user intentions.

\subsection{Recommendations for Design}
Based on our findings, we draw the following high-level recommendations for the design of AI-powered AAC interfaces to support the creation of timely humorous comments:

\textbf{Simplicity:} Participants valued simplicity as a key factor, particularly in two forms: minimizing the number of actions required to achieve their goal (generating a timely humorous comment) and offering an appropriate number of options.  Participants praised the \emph{Full-auto} interface for being able to quickly generate a joke and having an interface simple to understand. On the contrary, participants pointed out how the high number of steps required to use the \emph{Wizard} interface made it potentially unusable, despite offering one of the highest levels of agency to the user. Thus, we recommend future designers to carefully consider the number of “actions” the user needs to perform before they can get the AI to suggest humorous comments to use. We also recommend future designers to minimize the number of options that are simultaneously presented to the user. For example, future AI-powered apps could leverage the extended context window offered by more recent LLM models to retain information about users' preferences and patterns of use. Through this, the interface could adapt dynamically to offer the right amount of options (Low or high). This approach not only simplifies the UI but also reduces the overhead time that users spend reviewing multiple suggestions before deciding whether to use one suggestion or another.

\textbf{Recallability:} Participants pointed out that one of the best features of all our interfaces was having access to the full transcript of the conversation in the \emph{Context Bubble Selection} interface. The transcript not only allowed them to select the context bubbles for the AI to use but also as a source of material to think of jokes. Recalling P3 comments, being able to see the conversation while listening to their conversational partner helped her think of jokes and keep up with the conversation. Therefore, we recommend future designers to integrate a form of conversation visualization paired with an automatic speech recognition (ASR) toolkit. One way in which they could integrate such visualization could be following our example of the \emph{Context Bubble Selection} interface.

\textbf{Interoperability:} Even though all the users were able to interact with all the interfaces, we observed that P4 and P5 had a worse experience using their joystick input and eye gaze with some of the interfaces. Specifically, the action of scrolling was difficult to perform. As such, these participants did not find the interfaces requiring scrolling (\emph{Context Bubble Selection}, \emph{Wizard}) as helpful. The input experience of each individual is unique, and the effort to perform one task or another varies widely. We recommend future designers to (1) use non-scrollable pages to ease users with less mobility using alternative input methods and (2) offer multiple UI layout options since non-standard input devices like eye gaze keyboards might overlay another interface component on top of any application. This might obstruct the interaction of certain elements, as we observed with P5 and his Tobii Dynabox. 

\subsection{Future Work}

We explored four different interfaces during one study session, an important first step to get a sense of how to support AAC users in creating timely humorous comments during an ongoing conversation. Future work that conducts longitudinal studies outside the lab is needed to collect richer user interaction data as well as varied usage scenarios when generating timely humorous comments. Another avenue for future work is to conduct studies with a larger cohort of AAC users with a wider range of abilities. 

As stated in the introduction, there are many forms of humor besides the witty comments of this paper. More work is needed to support humor in the broad sense of the word. This would include studying other styles like sarcasm or irony, but also jokes that depend on deeper context between users than the current conversation, such as ``in-jokes''. With the advent of more powerful LLMs, and larger and multi-modal context windows, such projects become tractable.

Both our work and more expansive contextual AI systems do raise privacy concerns worth further studying. The reliance on AI for perpetual “listening” and transcription to generate jokes introduces privacy risks, potentially creating social discomfort and a fear of surveillance for both AAC users and those around them. Addressing these issues will require privacy-preserving designs that balance functionality with ethical considerations.

The reliance on proprietary and subscription-based AI services, such as Amazon Servers and GPT APIs, reduces accessibility and affordability. These recurring costs can be prohibitive for vulnerable users, reinforcing challenges like limited customizability and ecosystem lock-in. Future work should focus on open-source alternatives and cost-effective models to accelerate this shift and ensure equitable access to advanced AAC technologies.

Furthermore, our interviews found that the voice plays a central role in self-concept and identity~\cite{nathanson2017native}; future efforts could enable users to craft their own voices to reflect their unique personalities and nuance~\cite{pullin201517of} rather than defaulting to standardized tones. 

Finally, conducting this research while also being an AAC user provides valuable perspective, motivation, and connection to this work. These personal experiences motivated this work, and we plan to develop future auto-ethnographic and reflective research in the space of humor and AAC.

%% file: src/06_conclusion.tex
\section{Conclusion}

In this paper, we presented an exploration of how AI-powered interfaces can enhance Augmentative and Alternative Communication (AAC) technology to support AAC users in delivering well-timed humorous comments. Through qualitative interviews and a user study, we analyzed the challenges of using AAC for humor and proposed recommendations for designing interfaces that support humor creation with AAC. Future researchers can follow our recommendations for design to create intelligent AAC interfaces that provide AAC users with the ``freedom'' to express themselves and have a good laugh with others.